%% file: 3dLamp.tex
\documentclass{acmtog}

\acmVolume{VV}
\acmNumber{N}
\acmYear{YYYY}
\acmMonth{Month}
\acmArticleNum{XXX}

\usepackage{amsmath, amssymb}
\usepackage{url}
\usepackage{algorithm}
\usepackage{algorithmic}

\graphicspath{{figures/}}

\newcommand{\etal}{et al.}
\newcommand{\hidecomment}[1]{}

\providecommand{\shortcite}[1]{\cite{#1}}

\begin{document}

\markboth{H. Zhao et al.}{Printed Perforated Lampshades for Continuous Projective Images}
\title{Printed Perforated Lampshades for Continuous Projective Images}

\author{HAISEN ZHAO, LIN LU {\upshape and} YUAN WEI
\affil{Shandong University}
DANI LISCHINSKI
\affil{The Hebrew University of Jerusalem}
ANDREI SHARF
\affil{Ben-Gurion University}
DANIEL COHEN-OR
\affil{Tel Aviv University}
BAOQUAN CHEN
\affil{Shandong University}}




\maketitle

\begin{abstract}
We present a technique for designing 3D-printed perforated lampshades, which project  continuous grayscale images onto the surrounding walls. Given the geometry of the lampshade and a target grayscale image, our method computes a distribution of tiny holes over the shell, such that the combined footprints of the light emanating  through the holes form the target image on a nearby diffuse surface. Our objective is to approximate the continuous tones and the spatial detail of the target image, to the extent possible within the constraints of the fabrication process.

To ensure structural integrity, there are lower bounds on the thickness of the shell, the radii of the holes, and the minimal distances between adjacent holes. Thus, the holes are realized as thin tubes distributed over the lampshade surface. The amount of light passing through a single tube may be controlled by the tube's radius and by its direction (tilt angle). The core of our technique thus consists of  determining a suitable configuration of the tubes: their distribution across the relevant portion of the lampshade, as well as the parameters (radius, tilt angle) of each tube. This is achieved by computing a capacity-constrained Voronoi tessellation over a suitably defined density function, and embedding a tube inside the maximal inscribed circle of each tessellation cell. The density function for a particular target image is derived from a series of simulated images, each corresponding to a different uniform density tube pattern on the lampshade.
\end{abstract}

\input{intro}

\input{related}
\input{overview}

\input{results}
\input{conclusion}

\begin{acks}
This work is supported in part by grants from National 973 Program (2015CB352501), by NSFC (61232011, 61202147, 61332015), and by the Israel Science Foundation (ISF).
\end{acks}

\bibliographystyle{acmtog}
\bibliography{3dlamp}

\end{document}

%% file: intro.tex
\section{Introduction}
\label{sec:intro}

\begin{figure*}[!t]
\centering
\includegraphics[width=\linewidth]{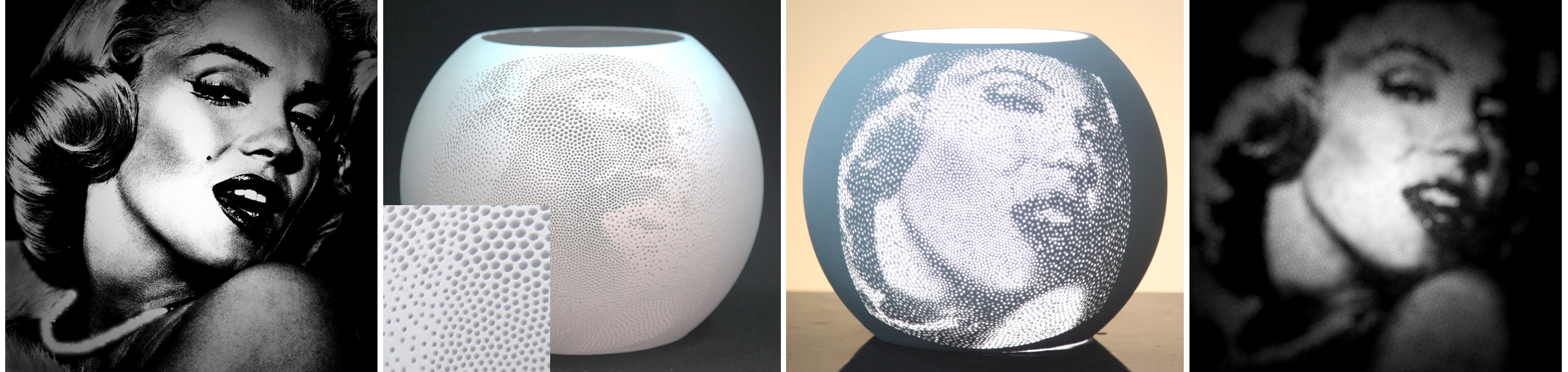}
\caption{A 3D-printed lampshade projecting a continuous grayscale image of Marilyn Monroe. From left to right: the target image, the 3D-printed perforated lampshade, the lampshade with a light source inside, and the resulting projection onto a nearby wall. The superposition of the individual light footprints on the wall forms a visually continuous image.}
\label{fig:teaser}
\end{figure*}

The emergence of 3D printing technology opens new interesting opportunities for
computerized halftoning. In this paper, we introduce the generation of
3D-printed perforated lampshades designed to project grayscale images onto
surrounding surfaces. The spatial distribution of tone across the projection is
generated by carefully controlling the amount of light shining through the
printed surface. Similarly to the dots used in halftoning, our basic idea is to
perforate the physical 3D surface of the lamp with small tubes, whose parameters
are used to control the amount and the spatial distribution of light that
emanates from the lamp and reaches a receiving surface. Figure~\ref{fig:teaser}
shows an example of such a lampshade and its projected image. Projecting continuous
tone imagery via a perforated surface can thus be regarded as \emph{halftoning with light} or \emph{3D halftoning}.

Halftoning is a technique used to represent continuous shades of gray through
the use of discrete dots of the same color, varying either in size, in shape, or
in spacing. The method relies on the natural low-pass spatial filtering of the
human visual system that blends the discrete dot pattern into a continuous tone.
Traditional halftoning uses simple-shaped dots, typically circular, of
continuously varying size. However, when the dots all have the same size,
spatial resolution can be traded for perceived tone
resolution~\cite{Ulichney1988,ModernHF08}.

The 3D halftoning technique that we present differs from ordinary halftoning in
a number of ways. First, unlike other common digital media, here one can
generate dots of continuous sizes. In that sense, the technique is closer to
analog halftoning, where the dot sizes are continuous. Second, dots are holes
in a surface, realized as tiny 3D tubes, hence having both \emph{radius} and
\emph{direction} with respect to the light source. Third, the resulting image is not
printed but projected, which requires to consider the geometry of both the
projecting surface (the lampshade) and the receiving surface (the wall). Lastly,
since the projected light footprint of each tube is slightly blurred, and
multiple footprints add up in areas of overlap, there is already some low-pass
filtering inherent in the image formation process. We explicitly account for and
take advantage of the effect of overlapping footprints of adjacent tubes, which
is inhibited in digital halftoning.

Moreover, it should be stressed that in our setting, the major challenge is to
ensure the printability and structural integrity of the perforated lampshade.
Specifically, there are strict lower bounds on the radii of the tubes, and on
the inter-tube spacing. Violating the former constraint would result in clogged
tubes, and the latter would render the shell fragile and prone to breakage. In
particular, having a lower bound on the tube radius means that darker tones
cannot be achieved by simply using smaller holes; instead, we reduce the amount
of light passing through a tube by tilting it away from the light source center.

We present a 3D halftoning technique that, given the geometry of the lampshade
surface, a target grayscale image, and a receiving surface, produces a spatial
distribution of tubes, along with their radii and directions (tilt angles), such
that the resulting projected image faithfully reproduces the input image. The
goal is to produce a continuous tone projected image, which strives to match the
distribution of tones and the spatial detail of the original.

Our approach attempts to match the target image intensity at each projected
location by placing tubes with suitable radii and tilt angles around the
corresponding location on the lampshade. Brighter areas are reproduced using
tubes with a larger radius, while in darker areas we place minimal radius
tubes, tilted away from the light source center. In order to maximize the
spatial resolution, the tubes must be placed as densely as possible, but they
must not violate the inter-tube distance constraint. To achieve this, we embed
each tube inside a disk that incorporates a safety margin around the tube. Note
that differently from ordinary halftoning or stippling, in our case both
brighter and darker areas require larger disks, with the maximal density of
disks corresponding to the middle of the grayscale range.

\begin{figure*}[ht]
\centering
\includegraphics[width=\linewidth]{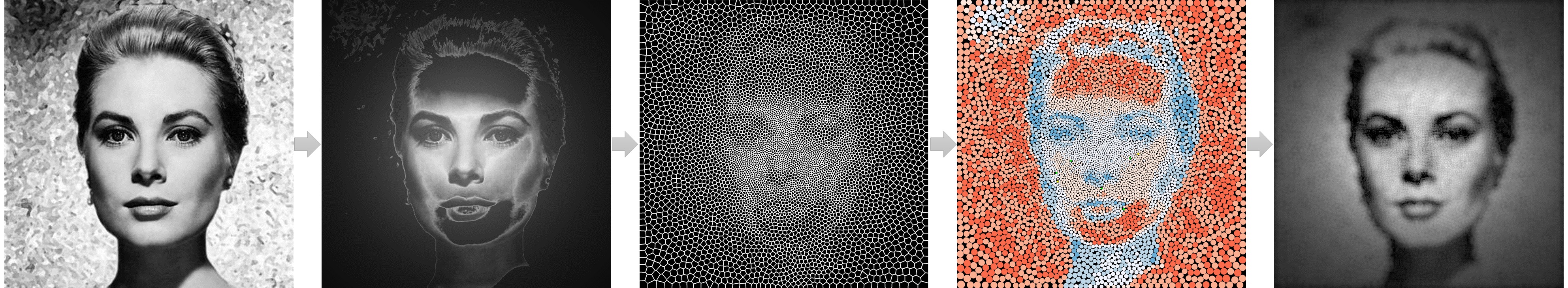}
\centerline{(a)\hspace{0.18\linewidth}(b)\hspace{0.18\linewidth}(c)\hspace{0.18\linewidth}(d)\hspace{0.18\linewidth}(e)}
\caption{Process overview: given a target image (a) we compute a density map (b) whose value at each location is inversely proportional to the required disk area.
Next, we compute a capacity-constrained Voronoi tessellation (c), and inscribe
a maximal disk inside each cell (d). Red shades indicate disks where the radius of
the embedded tube is larger than the minimal size (increasing the brightness),
while blue shades indicate disks where the embedded tube is tilted (decreasing
the brightness).
The simulation result of projected image on the receiving surface is shown in (e).
}
\label{fig:overview}
\end{figure*}

Having reduced the problem to one of finding a dense packing of disks with
spatially varying radii, we solve it by computing a capacity-constrained
Voronoi diagram over a suitably defined density function.
The density function for a particular target image is determined via lookup
from a series of precomputed simulated projected images, each corresponding
to a different uniform density tube pattern on the lampshade. The entire process
is illustrated in Figure \ref{fig:overview}.

In summary, the contributions of this paper are as follows:
\begin{itemize}
\item We tackle the novel problem of generating projected imagery by shining
      light through a 3D-printed perforated surface.
\item We introduce a novel 3D halftoning approach, where the halftoning dots are
      realized as a distribution of tubes with varying radii and orientation,
      passing through a solid surface.
      \item We present a method that determines the spatial distribution and the
      parameters of the individual tubes, while striving to match the target image,
      subject to fabrication constraints.
\end{itemize}
Our results demonstrate the effectiveness of the proposed technique using a
variety of target images. The limitations of the process are also demonstrated
and discussed.

%% file: related.tex
\section{Related Work}
\label{sec:related}

Halftoning is a classical technique which played a major role in traditional
paper printing and in digital displays~\cite{Kipphan}. Our work leverages this classical technique in the new domain of digital fabrication and 3D printing. Throughout the evolution of halftoning, the dominating issues have persistently been ones of structure preservation, tone reproduction, point density and spatial resolution.

\subsection{Halftoning and Stippling}

As a mature research problem, digital halftoning is well studied with numerous variants and solutions~\cite{Pang2008,SchmaltzGBW10}.
However, fewer researchers have addressed the problem of structure preserving halftoning and stippling.
Kim~\etal~\shortcite{Kim2008egsr} present a dot placement algorithm that guides the dot placement along feature flow extracted from the feature lines, resulting in a dot distribution that conforms to feature shapes.
Pang~\etal~\shortcite{Pang2008} introduce an optimization-based halftoning technique that preserves both the structure and tone metrics by optimizing for both features.
Chang~\etal~\shortcite{Chang2009} further extend the iterative optimization to an error diffusion method, which is more efficient.

Li~\etal~\shortcite{Li2010} show an efficient application for anisotropic blue noise sampling to stippling of images.
Li and Mould~\shortcite{Li2011} present an automatic structure-aware stippling algorithm. The method starts from a contrast-aware halftoning result and reduces the number of dots using a nonlinear priority-based error adjustment function.

Balzer~\etal~\shortcite{Balzer2009} propose using the capacity-constrained Voronoi
tessellation (CCVT) for enforcing the constraint of equal weighted area for the
region around each point in a stippled image. However, the method they propose
suffers from high computational complexity. Consequently, considerable attention
was given to developing faster alternatives. In particular, the remarkable work
of de Goes~\etal~\shortcite{deGoes2012} shows that CCVT can be formulated as a
constrained optimal transport problem. This results in a fast continuous
constrained minimization method, which is able to enforce the capacity
constraints exactly. In this work, we show how to encode the objectives and
the requirements of our 3D halftoning method as a density function, which enables
solving the problem by constructing a CCVT, and use the method of de Goes et
al.~for this purpose.

The works mentioned earlier are concerned with digital halftoning and stippling
of 2D images, while our work focuses on halftoning of 3D surfaces (with finite
thickness) in the context of 3D printing. In this context,
Stucki~\shortcite{Stucki1997} first introduced the idea of 3D digital halftoning
for transforming continuous-density objects into binary representations for
rendition with additive fabrication technologies.
Lou and Stucki~\shortcite{LouS98} further adapt the ordered-dither and error diffusion algorithms to the 3D case.
Zhou and Chen~\shortcite{Zhou093dHFT} utilize 3D digital halftoning, to reduce
the fabricating time in layered manufacturing.
In contrast to these works, we focus on 3D halftoning for the purpose of forming
projected imagery that reproduces the tones of a target image, subject to
fabrication constraints.

\subsection{Optics-related Fabrication}

Recently, design or generation of illumination effects via geometric modulation has been drawing increasing attention.
Mitra and Pauly~\shortcite{Mitra2009SA} introduce shadow art, an algorithm for computing a 3D volume, whose shadows best approximate multiple binary images.
Subsequent approaches generalize this to colored shadows using volumetric objects manufactured by transparent acrylic ~\cite{Wetzstein2011Layered3D,Baran2012}.
In general, they build transmittance functions to simulate light attenuation through multi-layered attenuators.

Our work is inspired by the technique of Alexa and Matusik~\shortcite{Alexa2012}, who drill holes with varying depths on a surface to induce the given image based on the occlusion of small holes.

Research in this area has also been addressing light reflections and refractions using surface modeling methods. These works produce a desired caustic image by casting a caustic, modulating the geometry using microfacets~\cite{Weyrich2009FMC}, micropatches~\cite{Papas11goal}, B-spline surfaces~\cite{Finckh2010}, continuous surfaces~\cite{Kiser2012,Yue2014} or normal fields~\cite{Schwartzburg2014}, and then milling or engraving the surfaces.

Researchers have also shown light effects in a wide range of applications, such as steganography and appearance design.
Papas~\etal~\shortcite{Papas2012MLR} develop passive display devices to hide images that can be decoded with fabricated lens.
Malzbender~\etal~\shortcite{Malzbender2012PRF} present a method for printing the 4D reflectance function of a surface for a fixed viewing direction.
Levin~\etal~\shortcite{Levin2013} use wave optics instead of geometric optics to produce custom designed reflectance functions with high spatial resolution.
Lan~\etal~\shortcite{Lan2013BAF} design the appearance fabrication by spatially-varying changes to both local shading frames and reflectance.

Printing optical fibers that control light propagation through total internal reflection between two surfaces has drawn some research attention thanks to the cutting edge multi-material 3D printers.
Willis~\etal~\shortcite{Willis2012} design optical fibers for customizing interactive devices in display and illumination.
Pereira~\etal~\shortcite{Pereira2014} introduce automatic fiber routing algorithm that minimizes curvature and compression for optimal light transmission.

All of the above are high-end techniques, which rely on expensive manufacturing equipment and specific materials.
In contrast, our method is designed for standard 3D printing, which is rapidly becoming an off-the-shelf, popular technology.
Furthermore, for widely used materials, such as plastic and powder, the transmittance of light is very weak, as light mostly scatters due to the surface roughness. Therefore, light effects based on surface reflection and refraction cannot be easily adapted to such materials.

%% file: overview.tex
\def \mx{\mathbf{x}}
\def \mc{\mathbf{c}}
\def \mr{\mathbf{r}}
\def \mX{\mathbf{X}}
\def \mV{\mathbf{V}}
\def \mp{\mathbf{p}}
\renewcommand{\labelenumi}{\alph{enumi})}

\newcommand{\rmin}{r_{\mathrm{min}}}
\newcommand{\dmin}{d_{\mathrm{min}}}
\newcommand{\rhomin}{\rho_{\mathrm{min}}}

\section{Perforated lampshade design}
\label{sec:method}

\begin{figure}[t]
\centering
\includegraphics[width=0.49\linewidth]{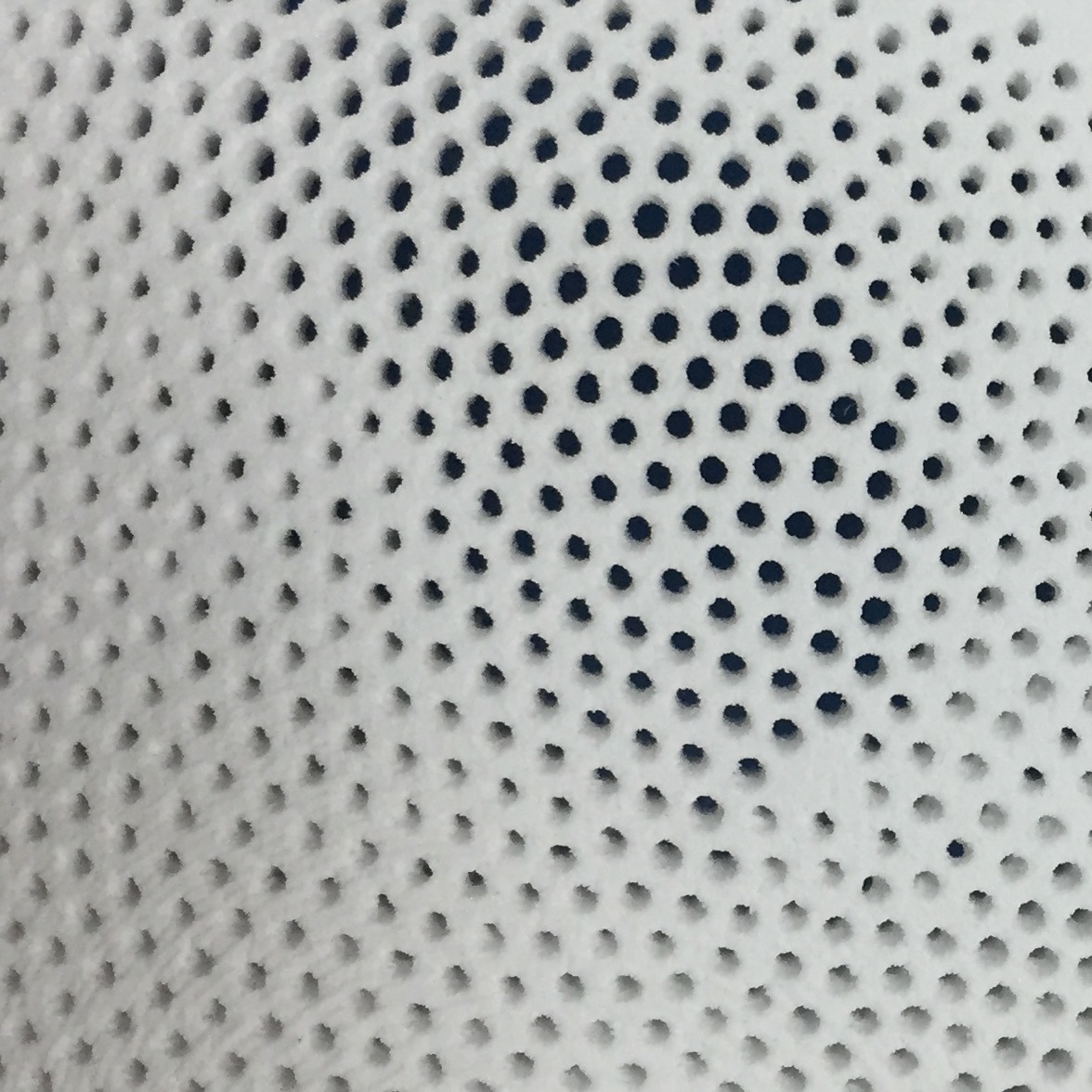}~
\includegraphics[width=0.49\linewidth]{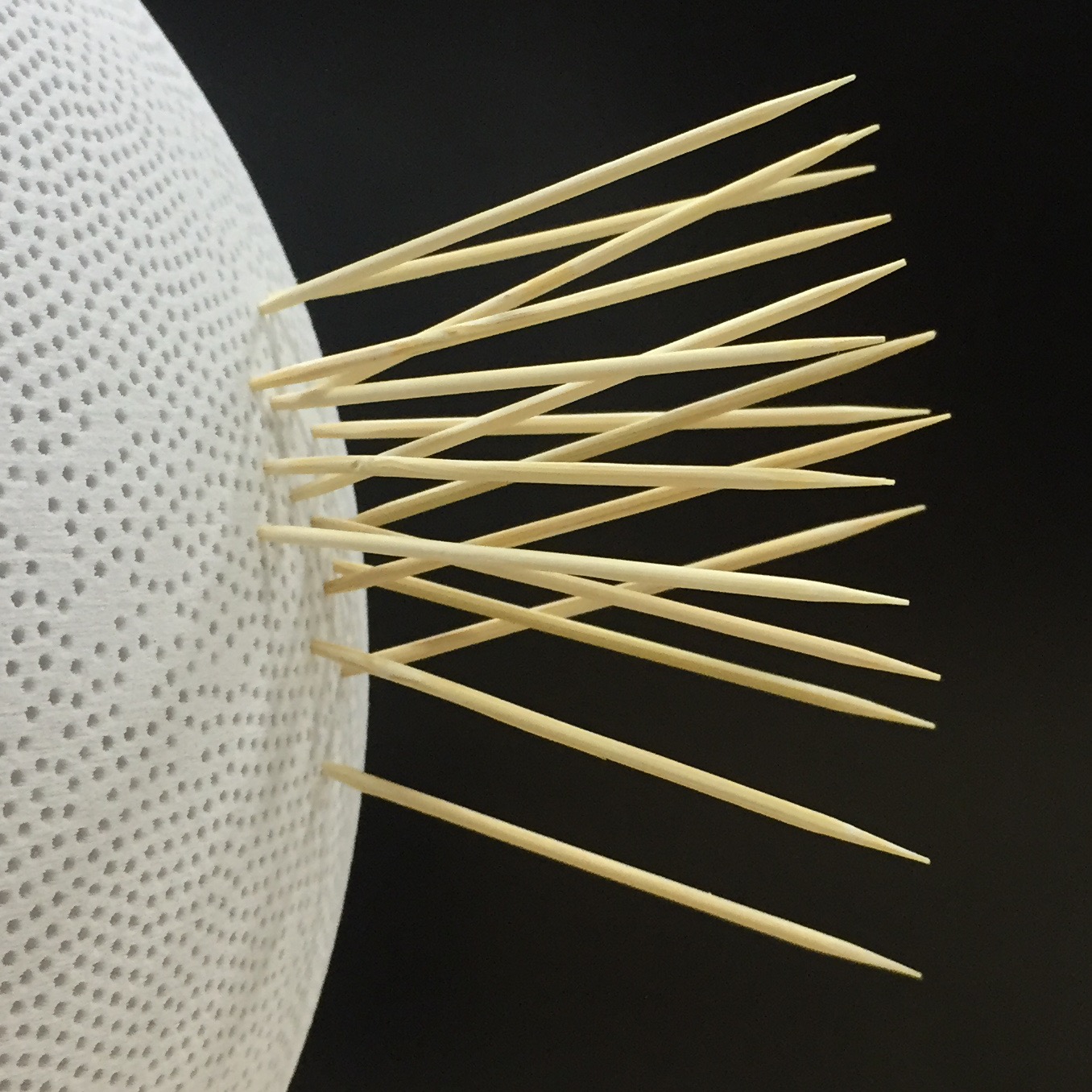}
\centerline{(a) \hspace{0.45\linewidth}(b)}
\caption{The physical 3D-printed lampshade shell perforated with tubes of varying sizes and density (a), and different tilt angles (b), as may be seen from the directions of the toothpicks.}
\label{fig:tubes}
\end{figure}

Recall that our goal is to design a 3D-printable perforated lampshade, such that
the light emanating from the lamp forms a continuous tone image, which is as
close as possible to a target grayscale input. Light emanates from the lamp
through holes in the lamp's surface. Due to the finite thickness of the lamp's
shell, the holes may be thought of as tiny tubes, whose density, radii, and
orientations vary across the surface, as may be seen in Figure \ref{fig:tubes}.
Since the total distribution of light across the projected image is determined
by the combined effect of light passing through the tubes, our challenge is to
determine the tubes' parameters, while respecting fabrication constraints imposed
by the need to obtain a printable and structurally sound surface.

Specifically, given a target grayscale image $I^t$ our task is to configure a set of tubes perforating the lampshade shell, such that the following objectives and requirements are satisfied:
\begin{enumerate}
  \item The projected image $I^p$ on a given diffuse surface, henceforth referred to as the \emph{wall}, closely approximates the target grayscale values across the projection. Note that the approximation is up to some scaling factor, since we have control over the total amount of luminous flux emitted by the light source, as well as over the exposure time, when capturing the projection with a camera.
  \item The projected image should exhibit continuous tones, while resolving fine
   spatial detail, as much as possible. In order to achieve this objective the
   density of the tubes should be maximized, while their radii should be minimized.
  \item Fabrication constraints: (a) the radius of a tube is bounded below by $\rmin$; and (b) any two adjacent tubes must have a gap of width greater than $\dmin$ of solid material between them. In our current setting, $\rmin = 0.6\mbox{mm}$ and $\dmin = 0.5\mbox{mm}$.
\end{enumerate}

\begin{figure}[t]
\centering
\includegraphics[width=.9\linewidth]{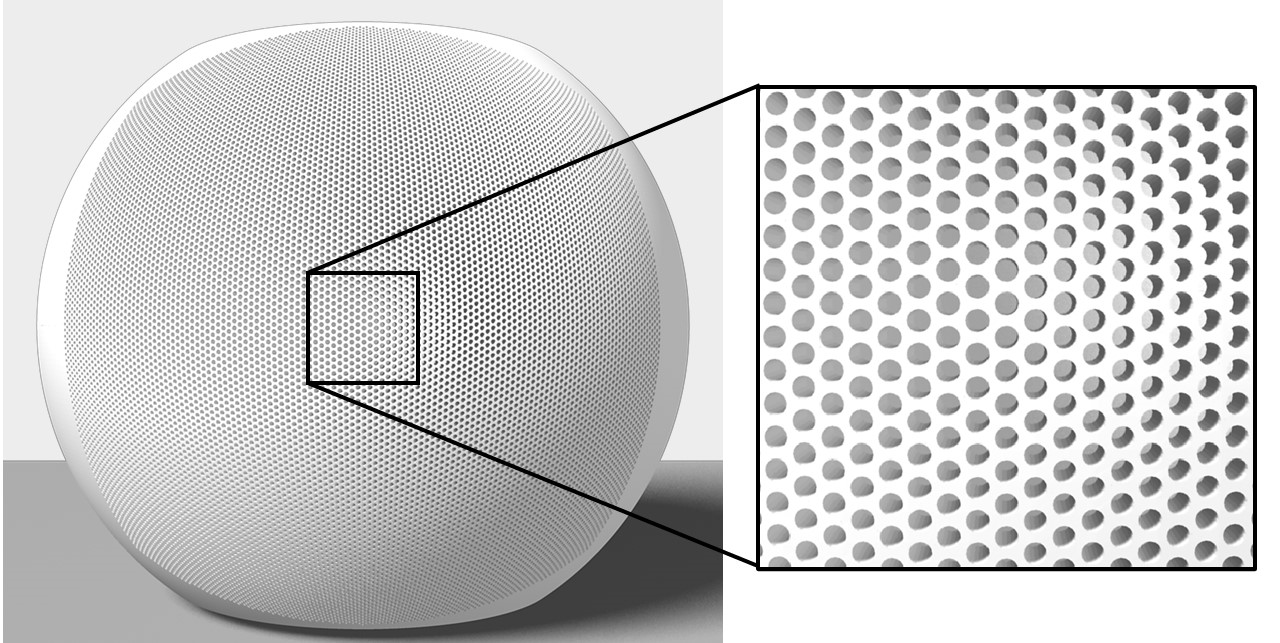}\\
\centerline{(a)}
\includegraphics[width=.45\linewidth]{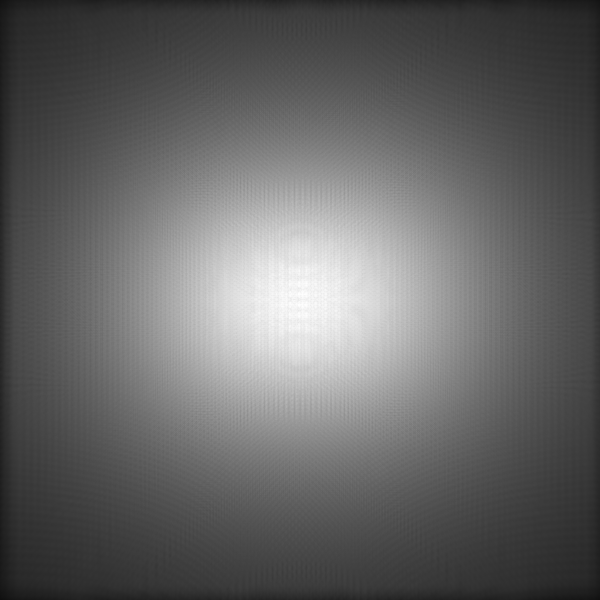}~~~
\includegraphics[width=.45\linewidth]{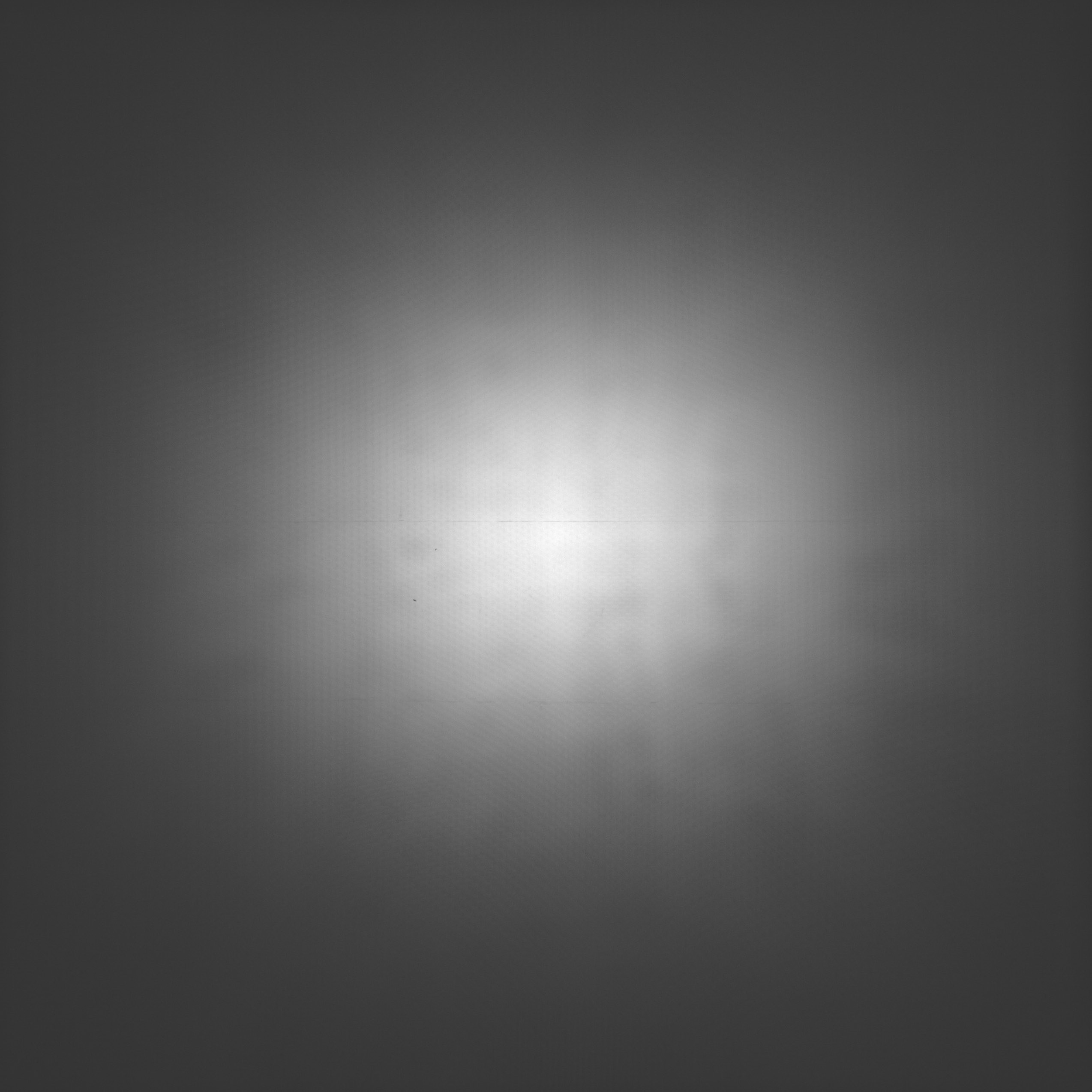}\\
\centerline{(b)\hspace{0.43\linewidth}(c)}
\caption{(a) Maximally dense packing of the smallest printable tubes, oriented towards the center of the light source. (b) The corresponding simulation result.
(c) A photograph of the actual projected image corresponding to this pattern.
}
\label{fig:init}
\end{figure}

\begin{figure}[t]
\centering
\includegraphics[width=0.49\linewidth]{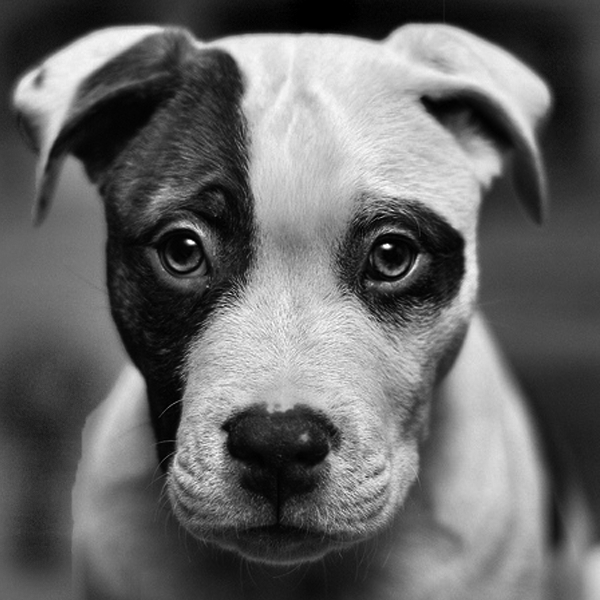}~
\includegraphics[width=0.49\linewidth]{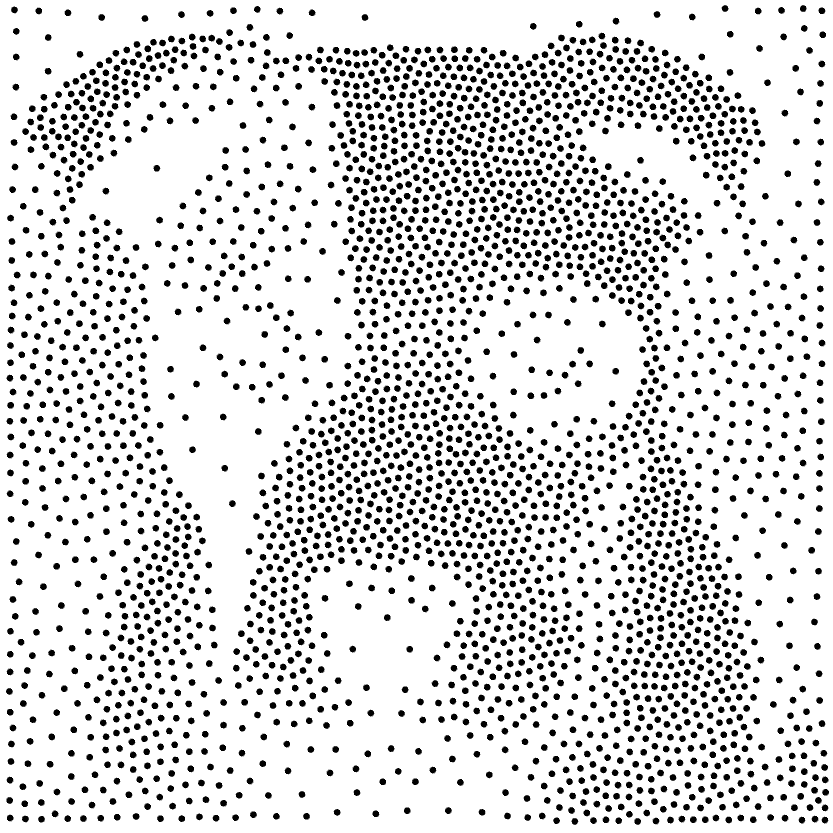}
\centerline{(a) \hspace{0.45\linewidth}(b)}
\includegraphics[width=0.49\linewidth]{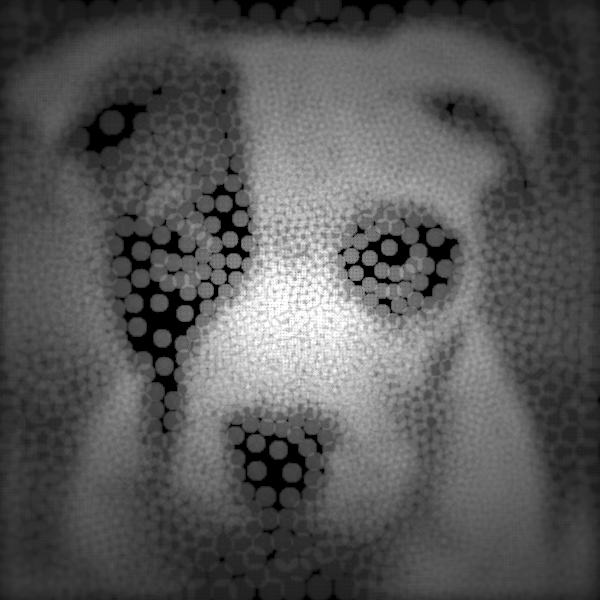}~
\includegraphics[width=0.49\linewidth]{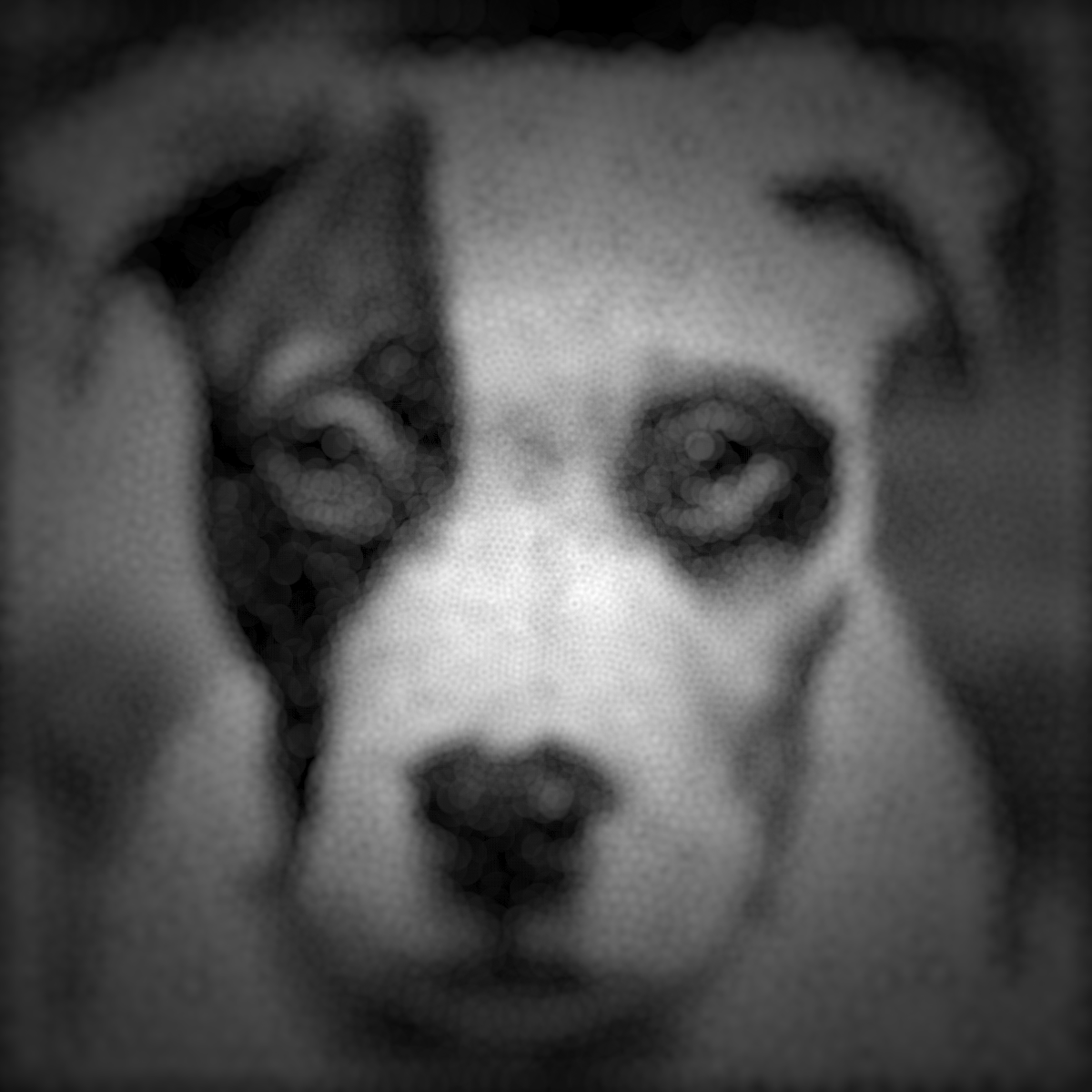}
\centerline{(c) \hspace{0.45\linewidth}(d)}
\caption{
For an input image (a), if we directly apply the stippling result (b) while ensuring the printability constraints, continuity and target grayscale tones cannot be achieved simultaneously in the projected imagery (c). Our method achieves a continuous result (d) by employing tube tilting to darken the tone.}
\label{fig:stippling}
\end{figure}

The densest arrangement of tubes that satisfies the fabrication constraints above
is shown in Figure \ref{fig:init}(a). This arrangement corresponds to a hexagonal
packing of disks of radius $\rmin+0.5\,\dmin$, with a tube of radius
$\rmin$ embedded at the center of each disk. The resulting projected pattern is
shown in Figure \ref{fig:init}(c). In order to match a target image $I^t$, the
tube arrangement must be adjusted in a spatially varying fashion.
In order to achieve brighter tones we must increase the tube radii,
but darker tones cannot be achieved by decreasing
the radii, since this would violate the first fabrication constraint above.

One alternative is to employ a stippling approach to generate a variable density set of tube positions that would
satisfy the fabrication constraints above. However, such an approach necessitates spacing tubes much farther
apart from each other in the darker areas of the image, resulting in projection of isolated light dots inside 
such areas. Thus, continuity and spatial detail resolution has to be severely compromised.
Instead, we propose decreasing the amount of light passing through tubes in darker areas by tilting their directions
away from the light source center. The tilting direction is random, in order to avoid undesirable structured artifacts.
This solution avoids unnecessarily isolating tubes, and plays an essential role in achieving continuous gray tones
in the projected imagery. 
It is worth mentioning that the tilting angle is not restricted by fabrication capabilities, which offers enough
variable tuning range.

The inadequacy of a stippling-based approach is demonstrated in Figure~\ref{fig:stippling}. Here, starting from
an input image (a), we use a state-of-the-art stippling method \cite{deGoes2012} to generate the pattern in (b).
In order to satisfy the fabrication constraints, the number of dots in this pattern is limited to roughly 3K. This results
in a (simulated) projected image with poor continuity and isolated light spots, especially inside the darker tone
areas around the dog's eyes and nose. Using our proposed solution, we are able to distribute roughly 6K tubes
across the lampshade, thereby achieving the same apparent grayscale range, but in a much more continuous
fashion.

Note that in order to ensure a minimal inter-tube distance, both widened tubes
and tilted tubes must be placed inside larger disks on the lampshade surface,
as illustrated in Figure \ref{fig:tubegeneration}.
The radius of each disk is chosen so as to accommodate the widened or tilted tube,
in addition to a \emph{safety margin} of width $0.5\,\dmin$. In summary, each
target intensity at a particular location corresponds to a certain disk radius,
and our goal is to distribute a set of tightly packed non-overlapping disks with
the desired radii across the relevant portion of the lampshade surface.

We attempt to achieve the required disk arrangement by computing a capacity-constrained Voronoi tessellation over a suitably defined density function, and inscribing a maximal circle inside each tessellation cell. The density function
for a particular target image is determined via lookup from a series of
precomputed simulated projected images, each corresponding to a different uniform density tube pattern on the lampshade; one such image is shown in Figure \ref{fig:init}(b).
The entire design process is visualized in Figure \ref{fig:overview} and
summarized in Algorithm \ref{alg:design}. The main stages of the process are described in more detail in the remainder of this section.

\begin{algorithm}[h!]
\caption{Perforated Lampshade Design}
\label{alg:design}
\renewcommand{\algorithmicrequire}{\textbf{Input:}}
\renewcommand{\algorithmicensure}{\textbf{Output:}}
\begin{algorithmic}

  \REQUIRE Grayscale image $I^t$, lampshade surface $S_L$, receiving wall surface
  $S_W$, light source
  \ENSURE Set of tube parameters (position, radius, tilt) on $S_L$

    \STATE
    \STATE \textbf{Preprocessing: simulate a series of reference images}
    \begin{itemize}
      \item Simulate the projected \emph{reference image} $B_0$, formed by
      tubes of radius $\rmin$, oriented at the light source center.
      Each tube is embedded in a disk of radius $\rmin+0.5\dmin$,
      with the disks arranged in a tight hexagonal grid on the relevant
      portion of $S_L$ (see Figure \ref{fig:init}).
      \item Simulate additional reference images $B_i$, $i = 1,\ldots,m$, formed
      by wider tubes of radius $\rmin + i\delta_r$. Each tube is embedded
      in a disk of radius $r_i = \rmin +|i\delta_r|+0.5\dmin$, arranged as above.
      We set $\delta_r = 0.05$mm, $m=10$.
      \item Simulate additional reference images $B_i$, $i = -1,\ldots,-m$,
      formed by a grid tilted tubes of radius $\rmin$.
      Each tube is oriented at a random
      direction away from the light source center, and tilted to the largest
      angle possible inside a disk of radius $r_i = \rmin + |i\delta_r| + 0.5\dmin$.
      The embedding disks are arranged as before.
    \end{itemize}

    \STATE \textbf{Compute the density function on $S_W$}
    \begin{itemize}
      \item For each image pixel at position $(x,y)$, find $k$ such that
      $B_k(x,y) \leq I^t(x,y) \leq B_{k+1}(x,y)$.
      \item Determine the desired disk radius $r_l$ on $S_L$ by linear
       interpolation between $r_k$ and $r_{k+1}$, the disk radii used
       by $B_k$ and $B_{k+1}$.
      \item Compute the corresponding disk radius $r_w$ on $S_W$.
      \item Set the density $\rho_w(x,y) \propto 1/r_w^2$, and normalize.
    \end{itemize}

    \STATE \textbf{Compute the disk distribution}
    \begin{itemize}
    \item Determine the desired number of disks $N$.
    \item Compute a capacity-constrained Voronoi tessellation using $N$
     disks and the density given by $\rho_w$.
    \item Inscribe the maximal circle inside each tessellation cell.
    \end{itemize}

    \STATE \textbf{Generate tubes}
    \begin{itemize}
    \item For each disk of radius $r$, centered at $(x,y)$ on $S_W$:
    If $I^t(x,y) \geq B_0(x,y)$, the tube radius is set to $r - 0.5\dmin$
    (the widest tube that fits inside the disk, with a safety margin, see
    Figure \ref{fig:tubegeneration}). The tube
    is oriented towards the center of the light source.
    \item If $I^t(x,y) < B_0(x,y)$, the tube remains with radius $\rmin$, and
    tilted away from the light source center (in a random direction),
    by the maximal angle afforded by a disk of radius $r - 0.5\dmin$ (see
    Figure \ref{fig:tubegeneration}).
    \end{itemize}

\end{algorithmic}
\end{algorithm}

\subsection{Simulating reference images}

We compute a series of simulated reference images $B_i$, each of which predicts the spatially varying reflected light intensity across the projection domain that results from a dense uniform arrangement of tubes with a given fixed radius and tilting angle. The reference images are precomputed and stored once for each setup (lampshade and wall geometry). The simulation process is described in detail in Section \ref{sec:simulation}.

Reference image $B_0$ corresponds to the densest arrangement of the smallest size tubes (radius $\rmin$) through the lampshade shell. The arrangement, the result of the simulation, and a photograph of the actual projected image are shown in Figure \ref{fig:init}. Images $B_i$ with $i > 0$ correspond to a similarly arranged grid of wider tubes, thus resulting in brighter images. Images $B_i$ with $i < 0$ correspond to grids of tilted tubes, resulting in darker images.

Given a target image $I^t$, these reference images enable us to determine the desired tube parameters at each position on the lampshade, in turn defining a density function on the wall. This density function is then used to compute a capacity-constrained Voronoi tessellation.

\subsection{Computing the density function}

We use the reference images $B_i$ to determine the spatially varying density
$\rho_w$ on the wall. Since the projected image is formed on the wall by a linear combination of light transmitted through the tubes, we first linearize the given target image $I^t$ by applying an inverse gamma correction. We assume that the target image is gamma corrected with a standard gamma value of 2.2.

For each projected image pixel $(x,y)$, whose target grayscale value is given by $I^t(x,y)$, we determine $k$, such that $B_k(x,y) \leq I^t(x,y) \leq B_{k+1}(x,y)$.
Let $r_k$ and $r_{k+1}$ denote the radii of the disks in which the tubes were embedded when generating $B_k$ and $B_{k+1}$. This means that the desired target
value at position $(x,y)$ could be achieved by embedding a tube inside a disk
of radius $r$, $r_k \leq r \leq r_{k+1}$, and we determine the value of $r$ using linear interpolation.

In our current implementation, we compute the capacity-constrained Voronoi
tessellation on the planar wall domain. Thus, we need to convert the desired
radii on the lampshade to ones on the wall. Specifically, given a radius $r$
of a disk on the lampshade, we compute the radius $r_w$ of a disk on the wall,
such that its projection back onto the lampshade (ellipse in the spherical case)
has the same area as the original disk of radius $r$.

Given the target disk radii $r_w$ across the projection area, we set the density to be proportional to the inverse of the disk area, $\rho_w(x,y) \propto 1/r_w^2$. The density values are normalized, such that $\max \rho_w = 1$. In practice, we
impose an upper bound of 1.3mm on the tube radius. Thus, the density function
is bounded from below, with the actual bound varying across the projection
surface.

\subsection{Computing the disk distribution}

Given the density function $\rho_w$, and a desired number of disks $N$, we use the method of de Goes et al.~\shortcite{deGoes2012} to compute a capacity-constrained Voronoi tessellation, and fit a maximal inscribed disk inside each of the resulting tessellation cells.

Although the above method is able to enforce the capacity constraints exactly,
the resulting Voronoi cells are not necessarily hexagonal, and thus some of the
maximal inscribed disks may deviate from their intended size. We estimate an
initial value for $N$ by computing $\rho_d$, the amount of density per disk
in the dense reference pattern $B_0$, and dividing the integral of the
target density function by $\rho_d$. We found that this estimate typically
results in a large percentage of disks being too small. We thus employ
a binary search to find the number of disks $N$ for which the percentage of
disks which achieve their intended radius (within 0.05mm) is greatest.

While most disks are acceptably close to their desired radius, there are
also disks which are too large or too small. Consequently,
the amount of light transmitted
through the embedded tubes could be smaller or larger than the desired intensity.
Larger disks are handled by reducing the embedded tube radius or tilt angle,
as needed to match the originally intended ones.
Smaller disks are resolved by removing a fraction of them.
For each group of such disks, whose Voronoi cells are connected, we remove disks randomly one by one.
After each disk removal, we enlarge the group of cells along the one-ring neighborhood to include their
immediately neighboring cells, and locally relax the remaining disks in
this area via centroidal Voronoi tessellation using Lloyd's method.

\begin{figure}[b]
\centering
\includegraphics[width=0.9\linewidth]{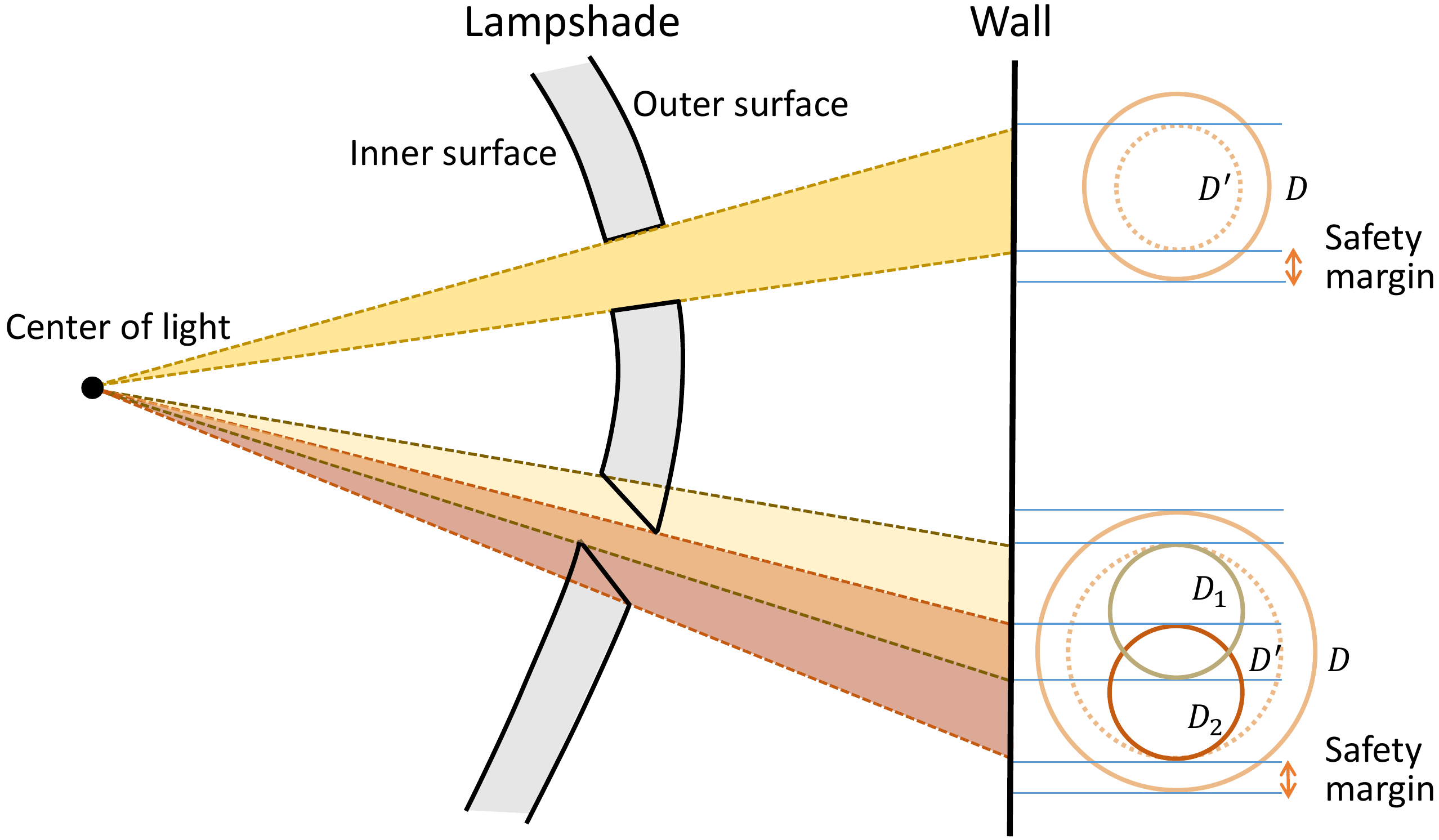}
\caption{Illustration of the tube generation.}
\label{fig:tubegeneration}
\end{figure}

\subsection{Tube generation}

For each disk $D$ centered at $(x, y)$ with radius $r_w$ on $S_W$, we first subtract
the safety margin to obtain a smaller concentric disk $D'$.
In the case where $I^t(x,y) \geq B_0(x,y)$, we generate a tube directed at the
center of the light source by intersecting the oblique circular cone defined by
$D'$ and the light source center with the lampshade shell, as shown in Figure
\ref{fig:tubegeneration}.

If $I^t(x,y) < B_0(x,y)$, a minimal radius tilted tube must be generated.
We select a random diameter of $D'$ and place two random minimal radius disks
$D_1$ and $D_2$ near the ends of this diameter. $D_1$ and $D_2$ are then
projected onto the outer and the inner surfaces of the lampshade shell,
respectively (see Figure \ref{fig:tubegeneration}).
The tilted tube is formed by connecting the two resulting contours.

At this point, a printable 3D model of the perforated lampshade may be generated
and  printed by a 3D printer.

\subsection{Projected image simulation}
\label{sec:simulation}

\begin{figure}[tb]
\centering
\includegraphics[width=0.9\linewidth]{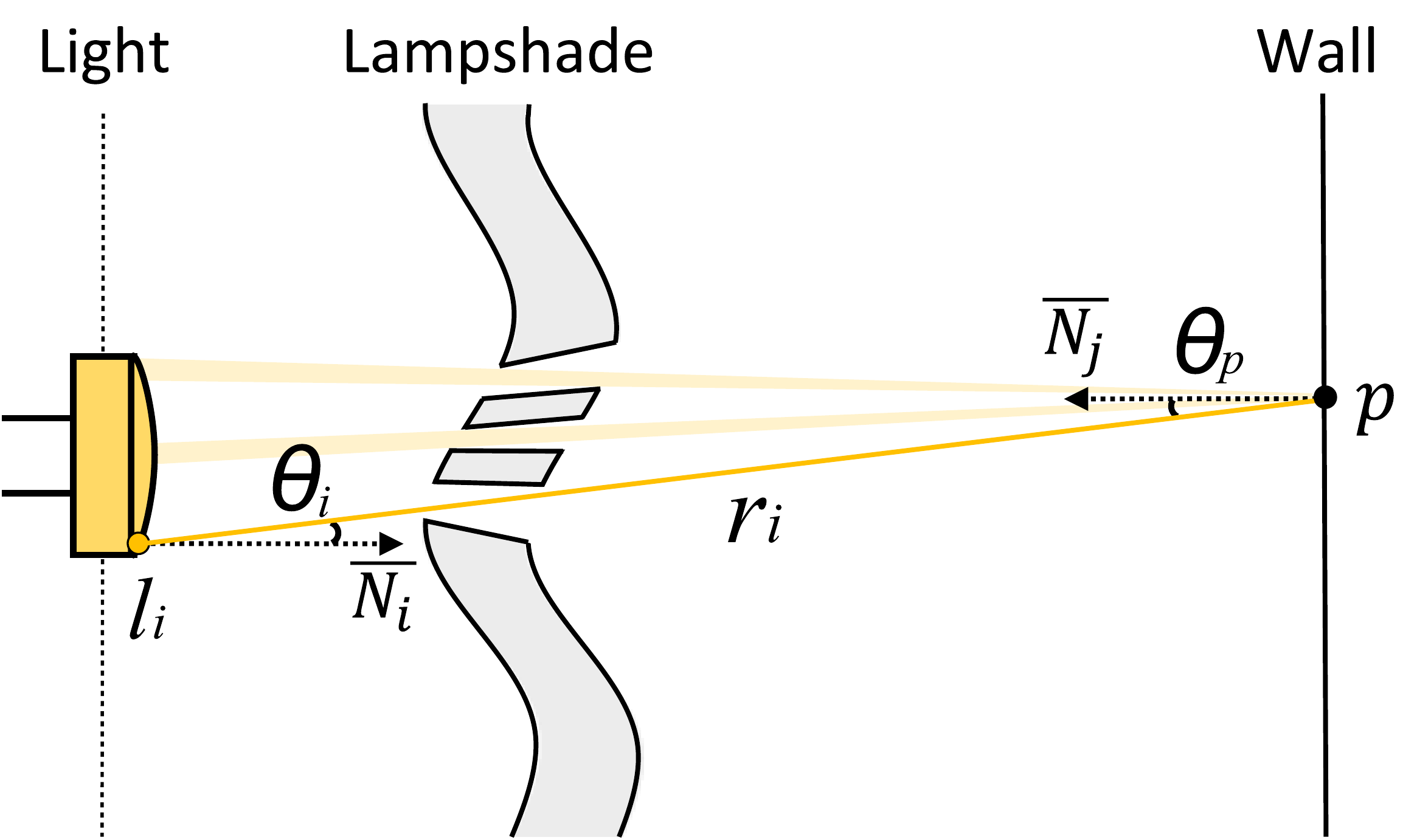}
\caption{Illustration of the illuminance computation.}
\label{fig:ff}
\end{figure}

The projected image is formed by the light emanating from the light source
inside the lamp, passing through the tubes in the lampshade and reaching a
 receiving surface (wall). Thus, the amount of reflected light at
each point on the wall depends on the emission characteristics of the light
source, the geometric relationship between the light source and the receiving
point, the visibility between them, and the reflectance characteristics of the
wall.

The light source that we use in this work is a small and powerful COB (Chips on
Board) LED, which is roughly disk-shaped with a diameter of 9mm.
In our simulations, we represent this area light source as a
collection of $n$ point light sources $\{l_i\}_{i=1}^n$, each emitting a
luminous flux of $\Phi_i=\Phi / n$, where $\Phi$ is the luminous flux for the
entire light source, measured in lumens. We use $n = 76$ in our results.

For a Lambertian light source and a Lambertian reflector, the total illuminance
$E_v(p)$ (in lux) reaching a point $p$ on the receiving surface is given by
\begin{equation}
  \label{eq:illum}
  E_v(p)=\Sigma_i \frac{\Phi_i}{\pi r_i^2}\cos(\theta_i)\cos(\theta_p)\,V\!(p,l_i),
\end{equation}
where $r_i$ is the distance from $p$ to $l_i$, and $\theta_i$ and $\theta_p$ are
the angles between the line connecting the two points and the two normals, as
shown in Figure \ref{fig:ff}. $V\!(p,l_i)$ is the visibility between
$p$ and $l_i$.

In practice, we found that our light source is not Lambertain, and its emitted
luminance diminishes as the angle $\theta_i$ increases. Furthermore, we found
that this behavior is slightly anisotropic, and the luminance diminishes more
quickly in the vertical direction. Empirically, we found that the non-Lambertian
behavior and the anisotropy are still well modeled by Equation (\ref{eq:illum}),
provided that the horizontal and vertical coordinates of the point $p$ (with respect
to the center of the image) are each scaled by an appropriate factor. Specifically,
the scaling factors that we use are 1.7 and 1.9 for the horizontal and the
vertical coordinate, respectively.

\begin{figure}[tb]
\centering
    \includegraphics[width=0.495\columnwidth]{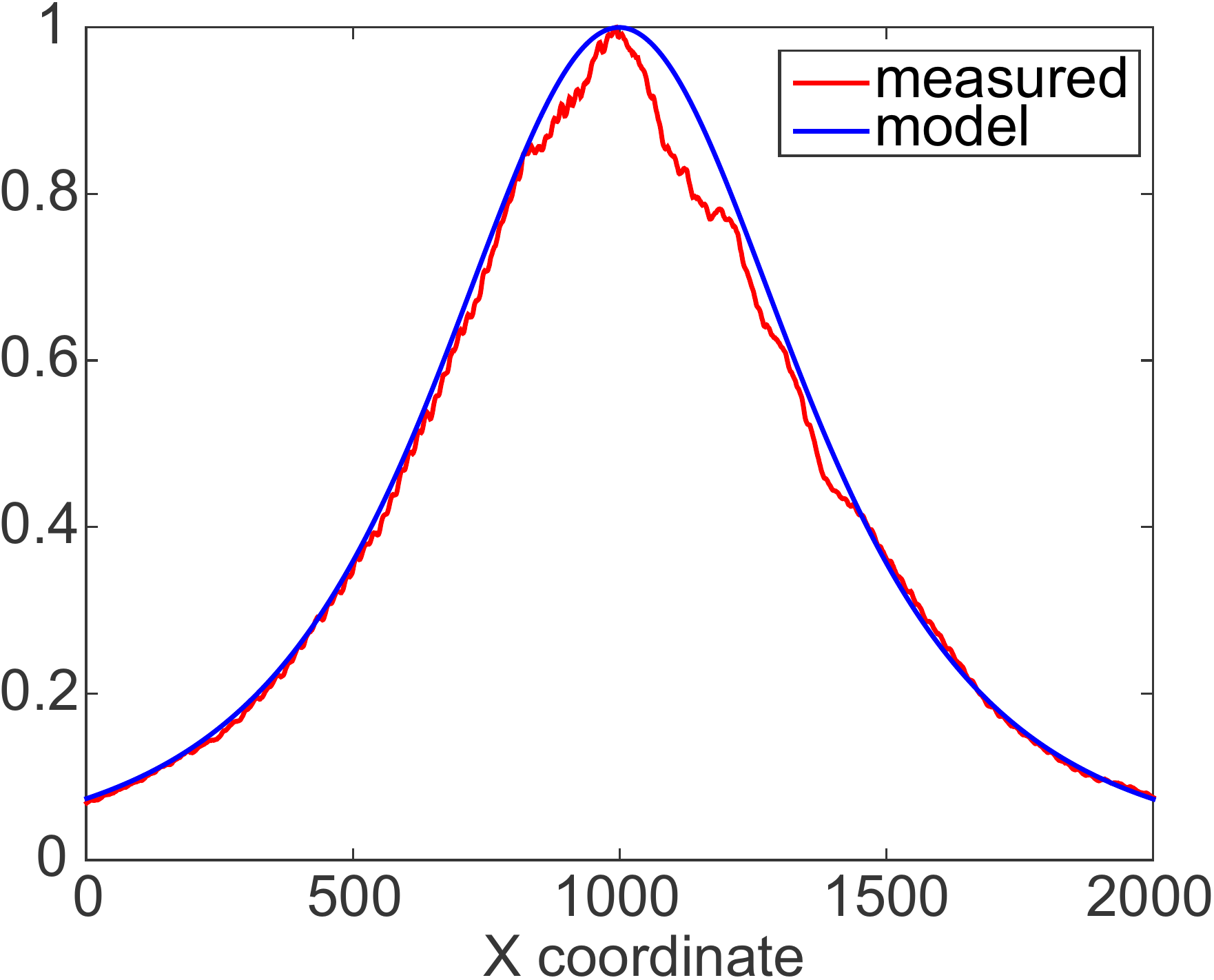}
    \includegraphics[width=0.495\columnwidth]{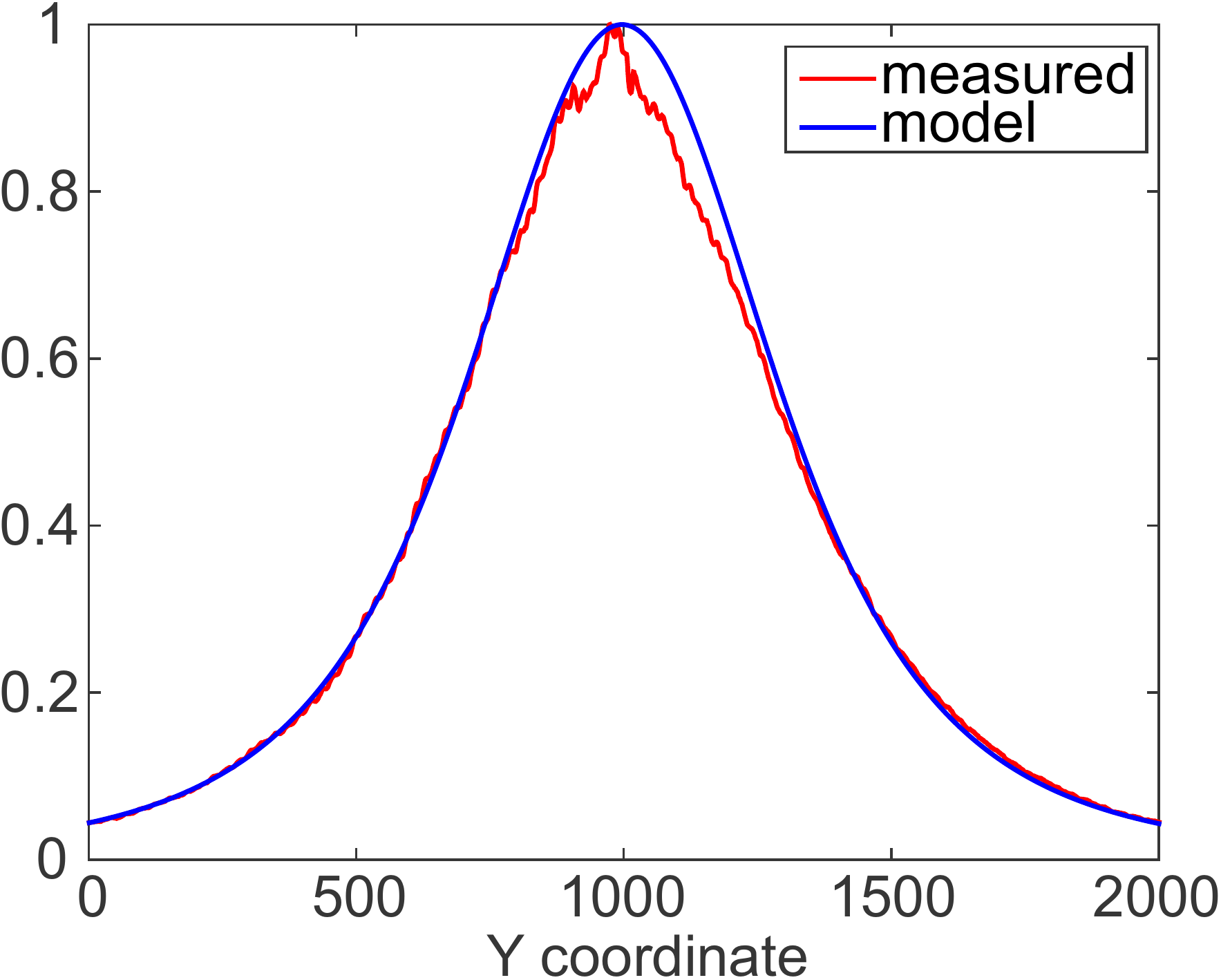}
\caption{The measured intensity of the projected image (red) is well predicted by our model (blue). Left: a horizontal slice through the center, right: a vertical one. Note that the horizontal and vertical falloff rates is different. }
\label{fig:modelfit}
\end{figure}

The accuracy of our empirical model is demonstrated in Figure
\ref{fig:modelfit}. The red curves show a horizontal (left) and a vertical (right)
cross section through the projected image shown in Figure \ref{fig:init}(c),
and the blue curves are the values predicted by our model.

We evaluate the visibility of points on the light source using ray casting (see
Figure \ref{fig:ff}). Note that it is not necessary to intersect rays with the
complete geometric model of the perforated lampshade shell. Instead, we quickly
determine a small set of tubes that are relevant for a given ray, and check
whether the ray passes through one of these tubes, without intersecting the tube's
surface. Note that the above approach accounts only for direct illumination,
without taking into account reflections, subsurface scattering, etc. In order
to assess the magnitude of these effects, we have sprayed the interior of
lamps with black paint, and found the differences in the measured intensity
very small. The main difference is that this reduces the amount of ambient
light in the room, leading to projected images with slightly better contrast.

\begin{figure}[tb]
\centering

\includegraphics[width=.102\linewidth]{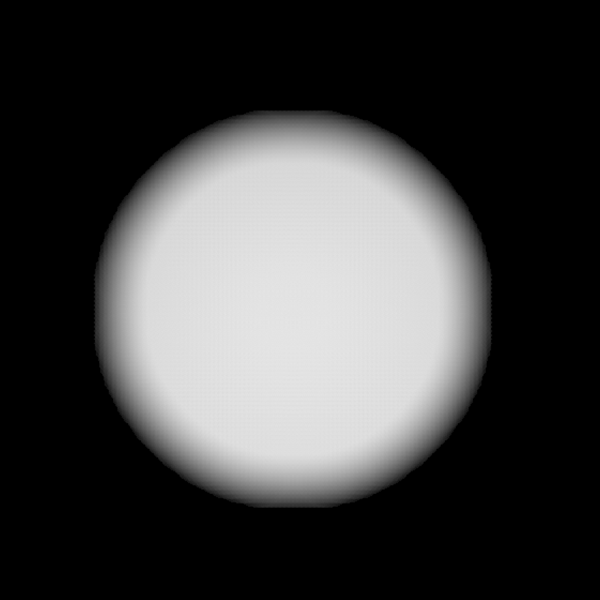}
\includegraphics[width=.102\linewidth]{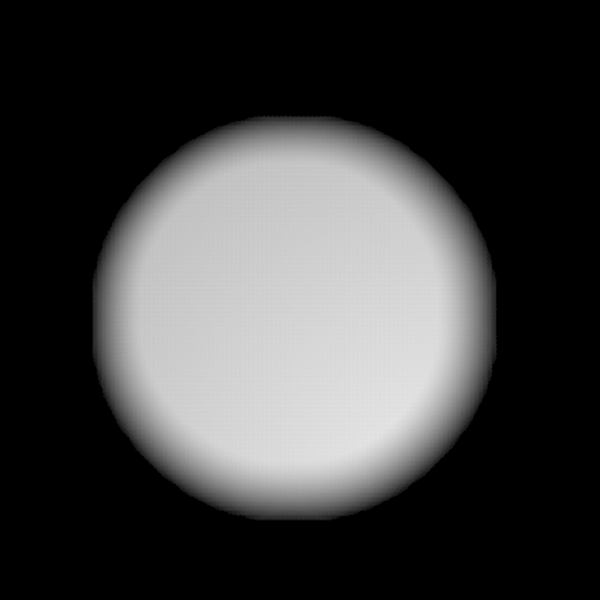}
\includegraphics[width=.102\linewidth]{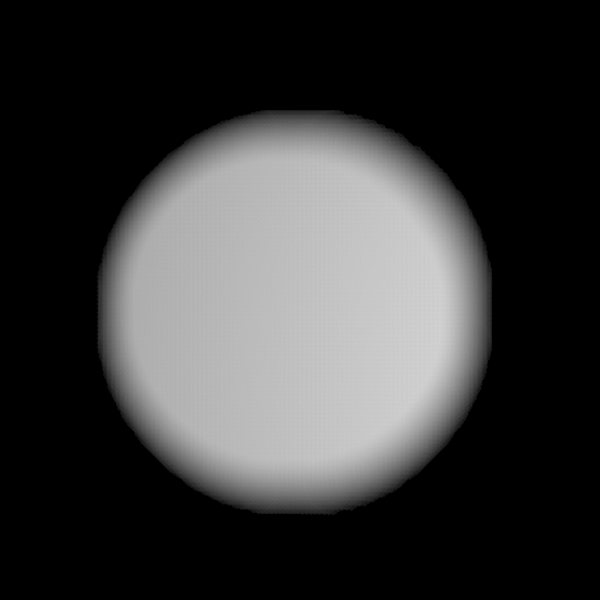}
\includegraphics[width=.102\linewidth]{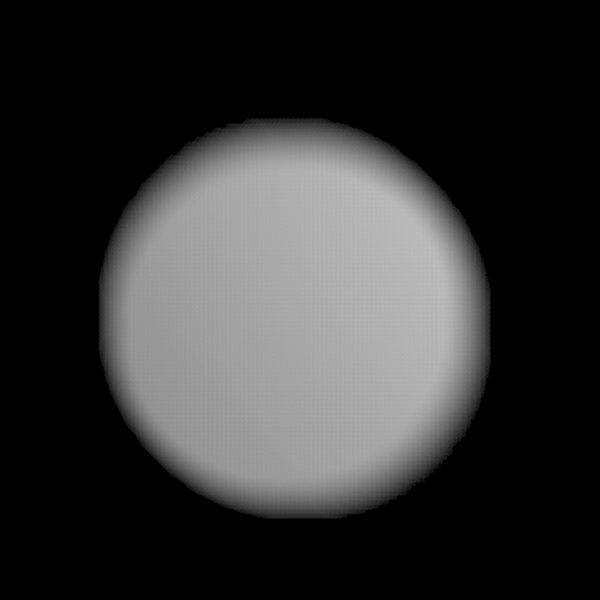}
\includegraphics[width=.102\linewidth]{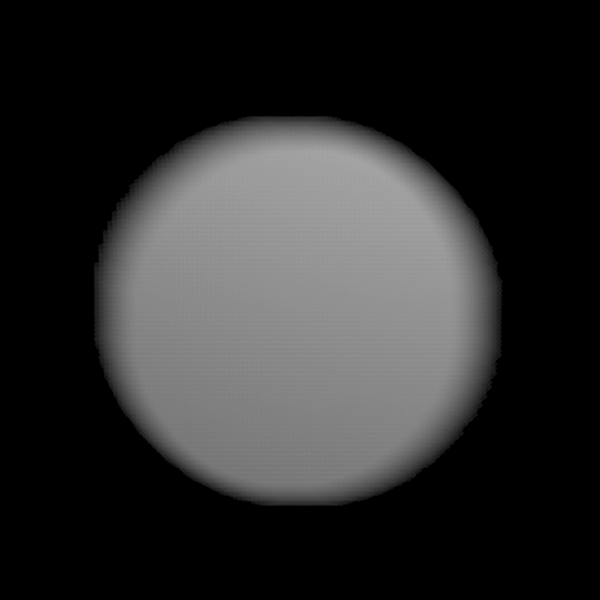}
\includegraphics[width=.102\linewidth]{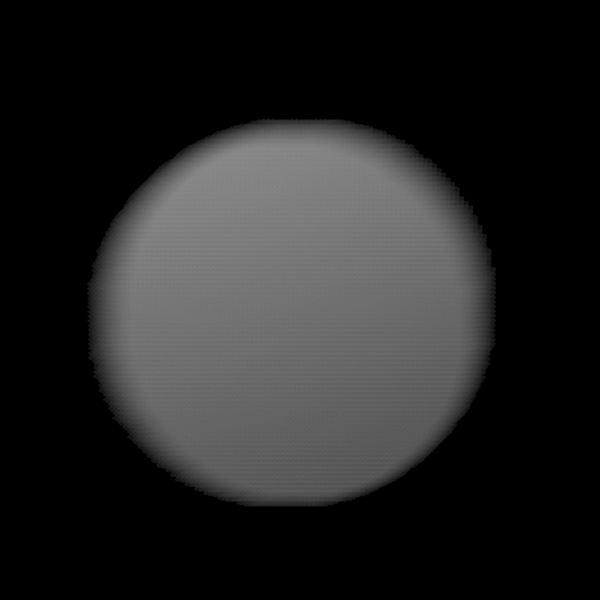}
\includegraphics[width=.102\linewidth]{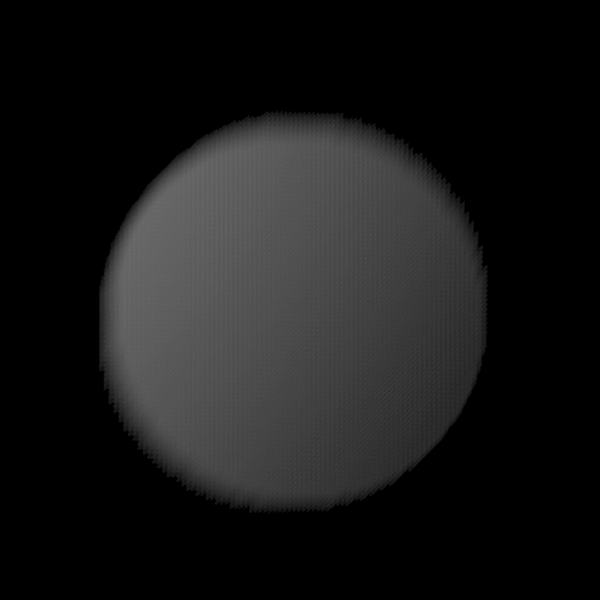}
\includegraphics[width=.102\linewidth]{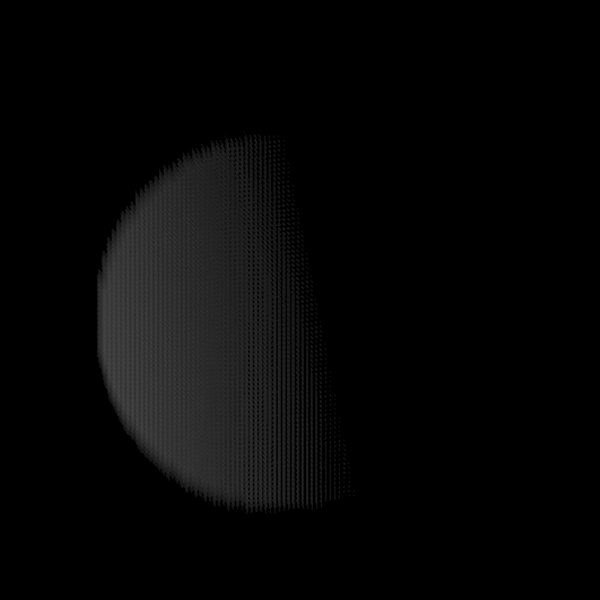}
\includegraphics[width=.102\linewidth]{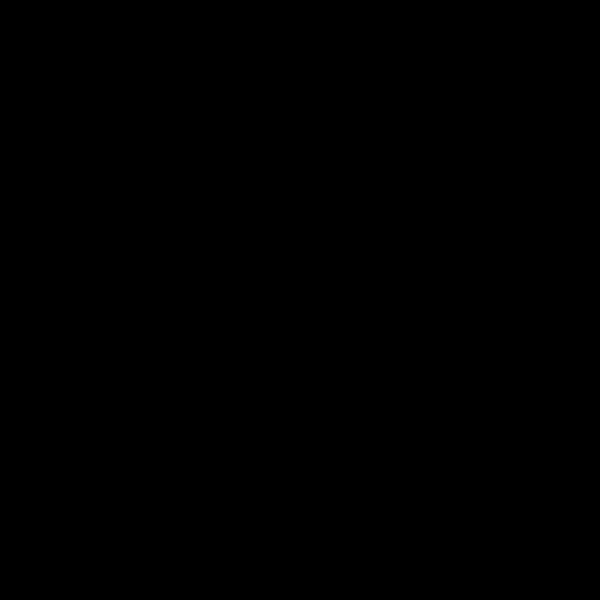}\\
 \vspace{3pt}
 \includegraphics[width=.102\linewidth]{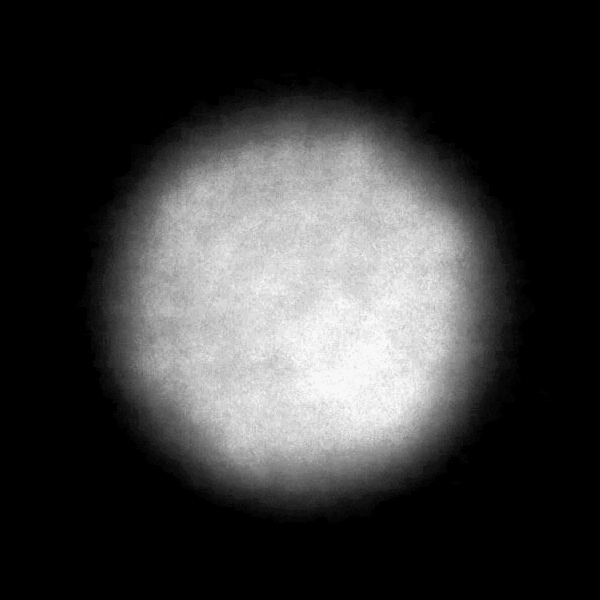}
\includegraphics[width=.102\linewidth]{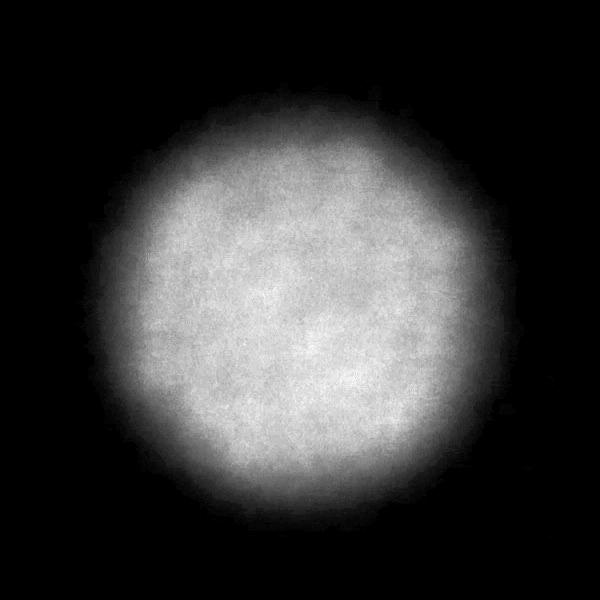}
\includegraphics[width=.102\linewidth]{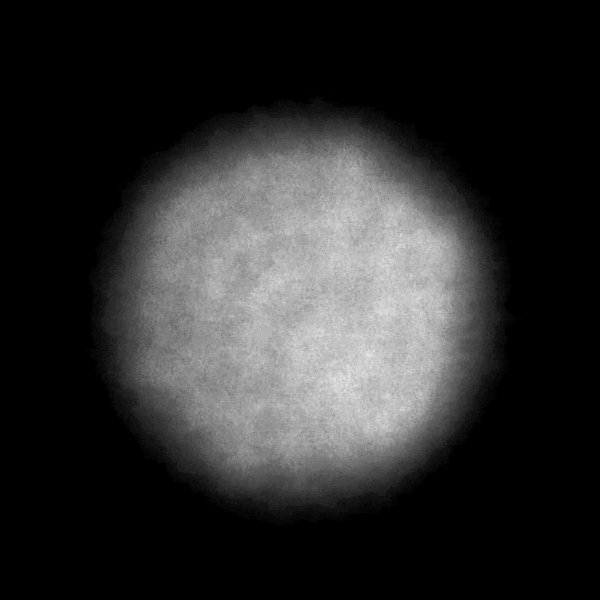}
\includegraphics[width=.102\linewidth]{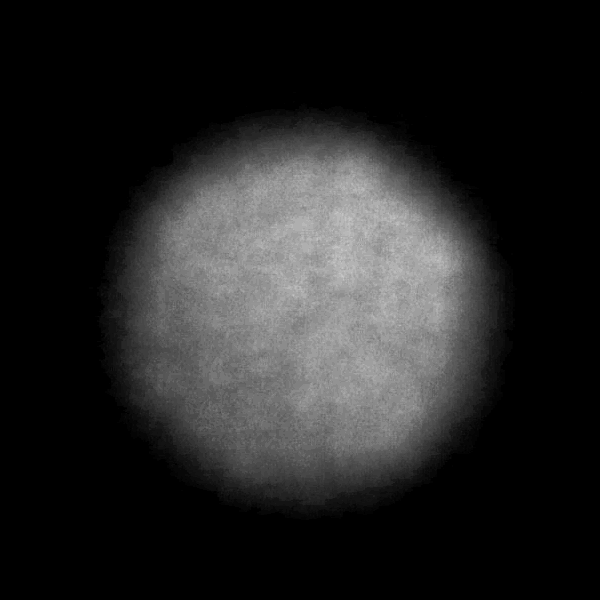}
\includegraphics[width=.102\linewidth]{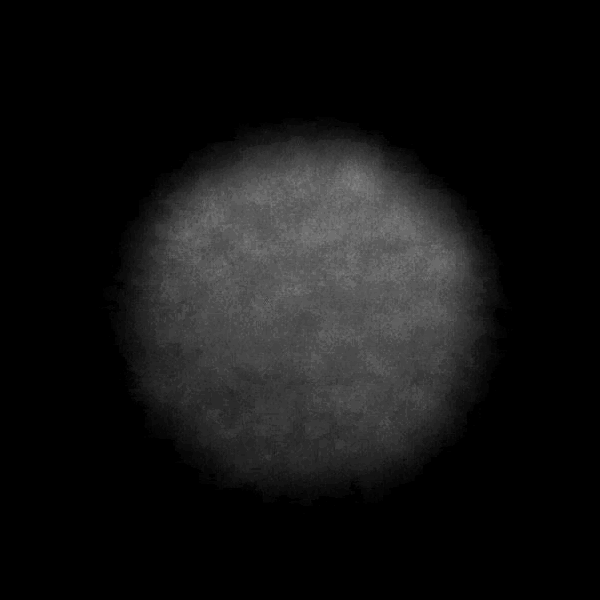}
\includegraphics[width=.102\linewidth]{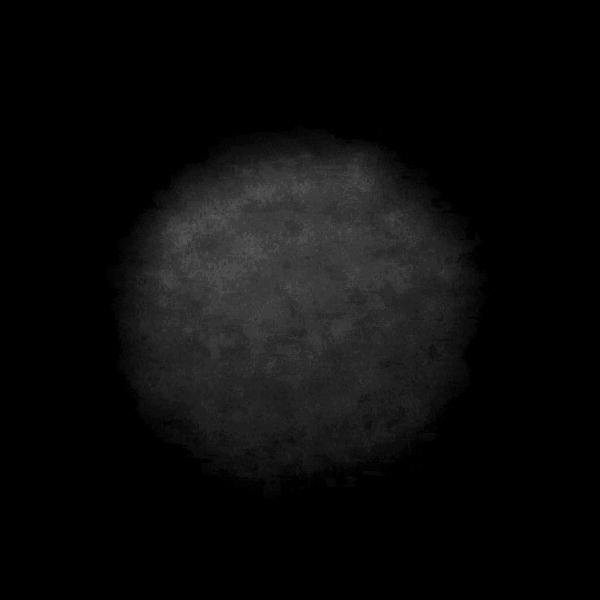}
\includegraphics[width=.102\linewidth]{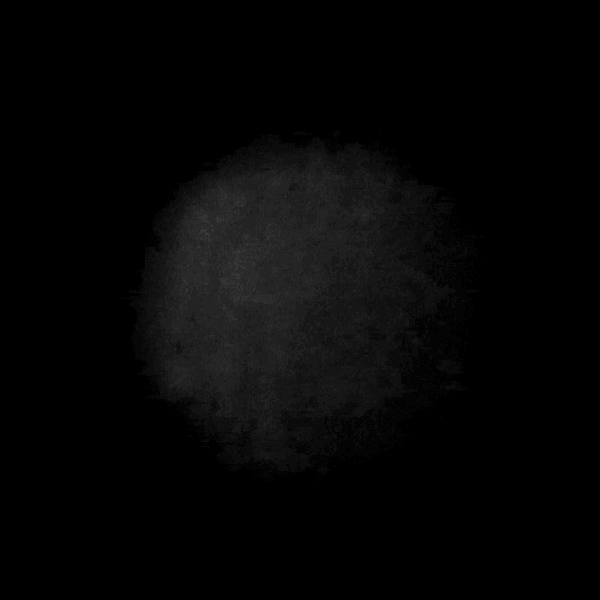}
\includegraphics[width=.102\linewidth]{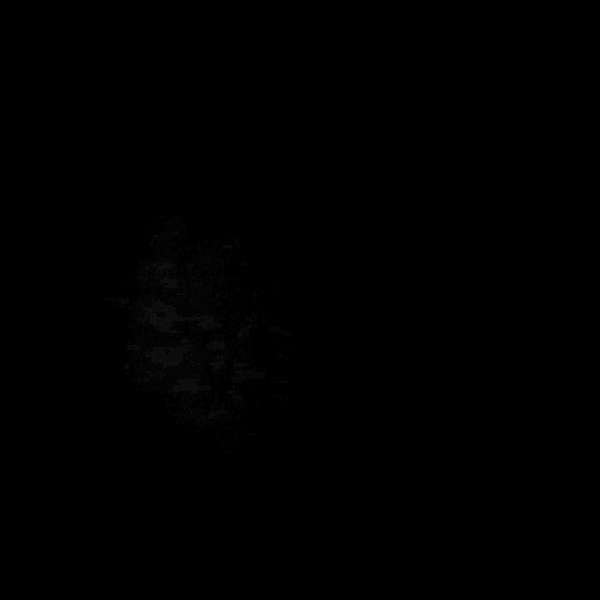}
\includegraphics[width=.102\linewidth]{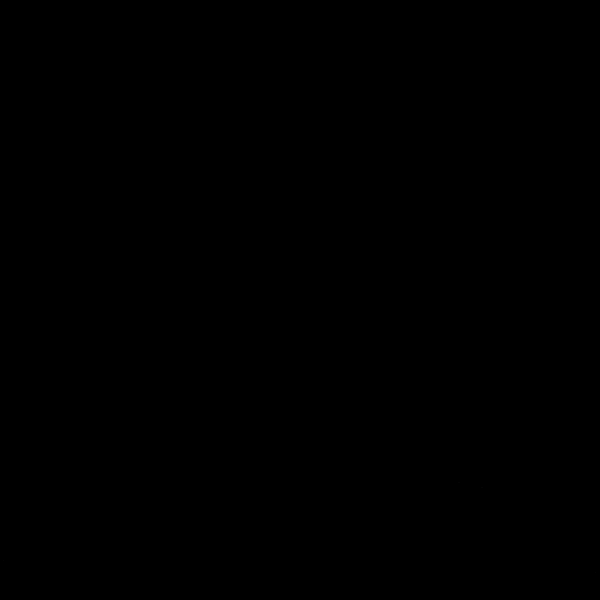}\\
\includegraphics[width=\linewidth]{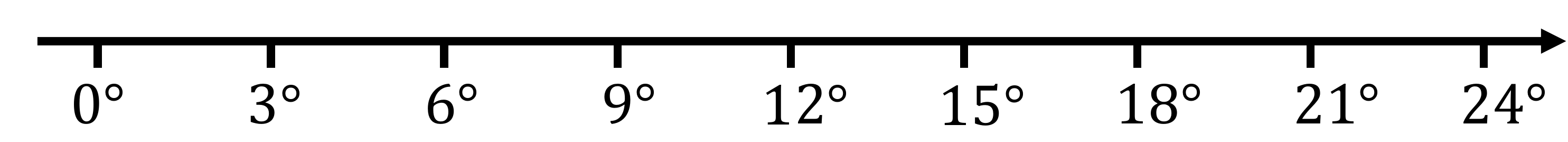}\\
\includegraphics[width=.6\linewidth]{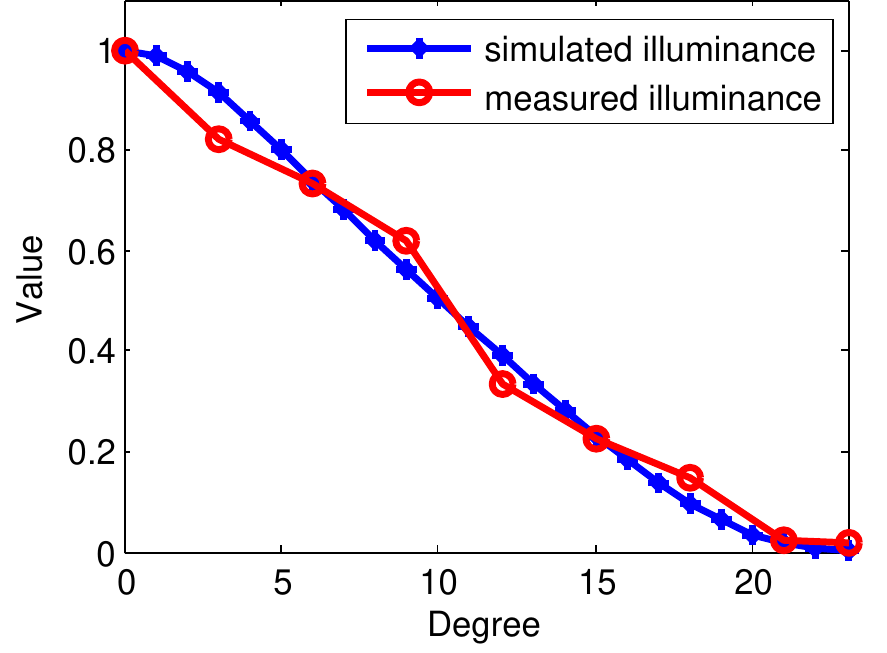}
\caption{One tube (diameter 1.2mm) with varying tilting angles. Top row: simulated footprints; Second row: the physical footprints;
Bottom: plot of the normalized sum of simulated illuminance values in the footprint and the sum of grayscales across an actual footprint photo.}
\label{fig:onetube}
\end{figure}

Accurate light source visibility estimation is particularly important for
correctly simulating tilted tubes. Figure \ref{fig:onetube} shows the simulated
and the actual light footprints of a single tube, for a variety of different tilt angles. The sum of the intensity
across the area of the footprint is plotted below (normalized such that the
maximal sum is 1.0). We can observe that the simulation predicts the
illuminance reasonably well, with some deviations between the two curves
due to imprecisions in the manufacturing process.

Finally, in order to visualize the simulated results we convert them into
grayscale values by scaling the result into a $[0,1]$ range, and applying a
gamma correction (we use gamma 2.2 in all our results).

%% file: results.tex
\section{Results}
\label{sec:results}

\begin{figure}[htb]
\centering
\includegraphics[width=.9\linewidth]{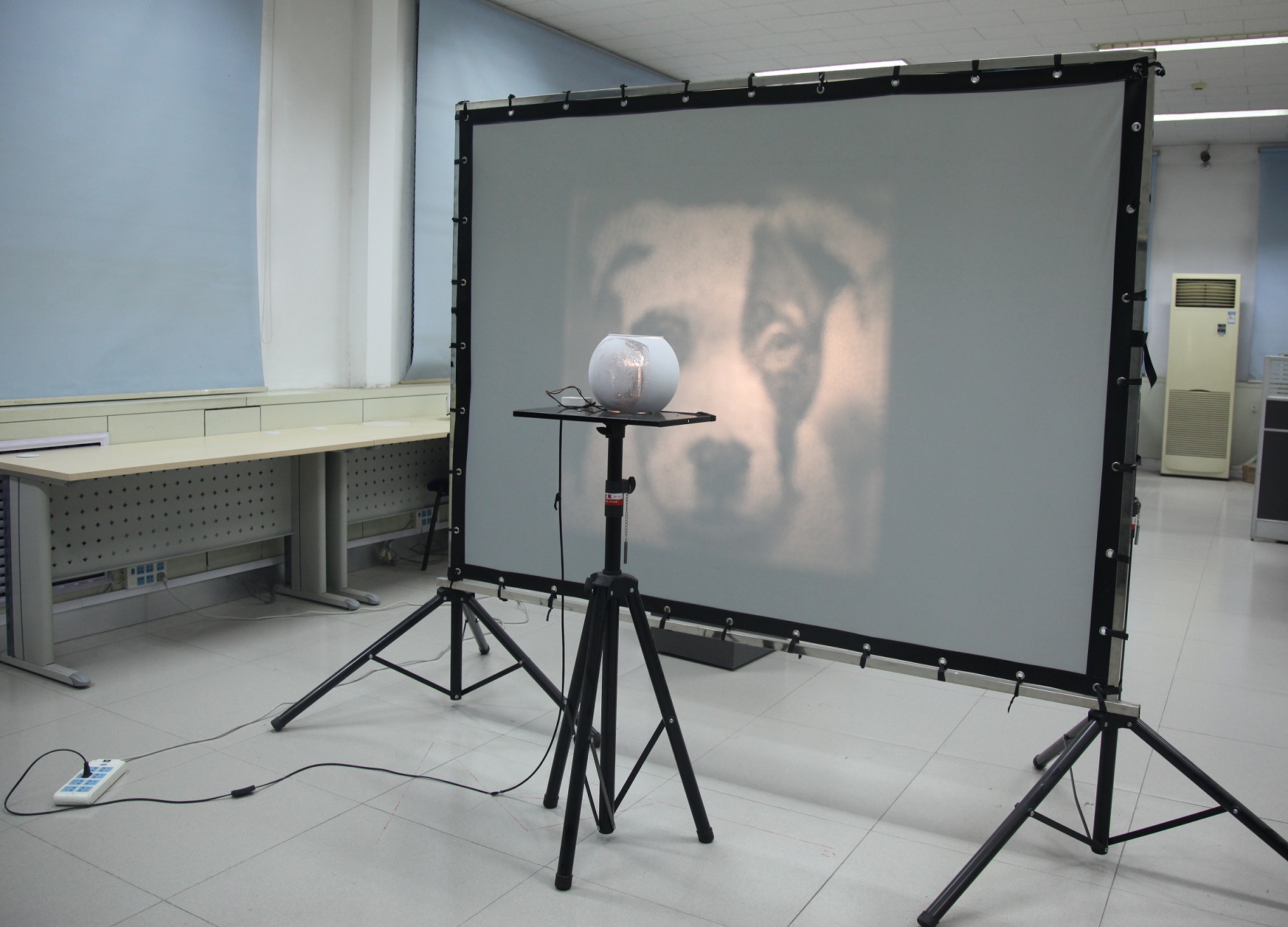}
\caption{A photograph of our setup.}
\label{fig:setup}
\end{figure}

\begin{figure*}[htb]
\centering
\includegraphics[width=\linewidth]{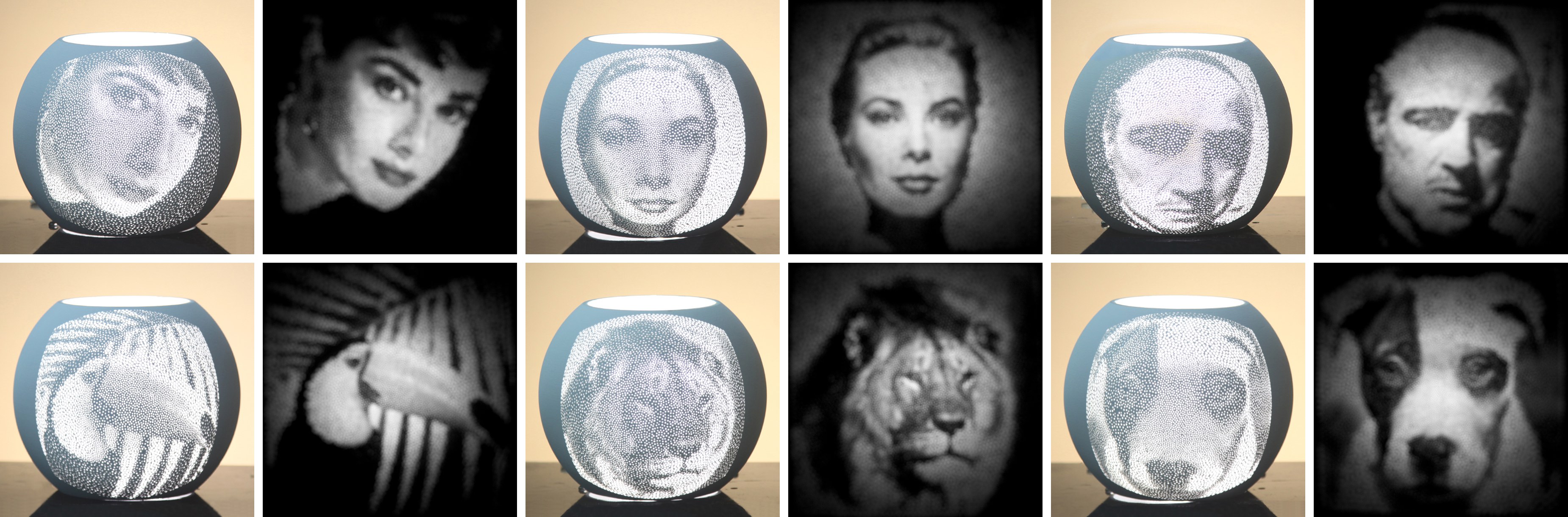}
\caption{Additional results on spherical lampshades. Left: our 3D-printed lampshade, with light projecting through the surface. Right: the photo of projected image. The lamps are 22cm in diameter.
}
\label{fig:spherical-results}
\end{figure*}

We have implemented the method described in the previous section,
and have used the resulting system to design a number of
lampshades for projecting a variety of target images. The results may
be seen in Figures~\ref{fig:teaser} and \ref{fig:spherical-results}.
The lampshades in all of these results have spherical geometry,
with a diameter of 22cm. To save material we
typically perforate each spherical lamp with two images on two opposing sides.
As mentioned earlier, the light source we use is a Cree$^\circledR$ XLamp$^\circledR$ CXA1507 LED with 3000K color temperature,
which is roughly disk-shaped with a radius of 9mm.
The perforated lamp projects its image onto a planar rear projection screen that offers high light transmission, located at a distance of 40cm from the light source.
The physical size of the projected image is 100 $\times$ 100
cm. Figure \ref{fig:setup} shows a photograph of our setup.
The photographs of the projected images shown in this paper are shot from the opposite side of the screen for eliminating distortion, using a Canon EOS 5D Mark \uppercase\expandafter{\romannumeral2}, with
exposure 1/80 sec, f/4.0, and ISO 400 settings, in a room without any
additional light sources.

\begin{table}[h]
\centering
\begin{tabular}{|c||c|c|c|}
\hline
Model        & Initial \#disks  & Final \#disks & \% correct size \\ \hline\hline
Marilyn      & 7323  & 5789    & 79.47   \\ \hline
CircularRamp & 7914  & 6321    & 87.73   \\ \hline
Hepburn      & 7617  & 5945    & 82.50   \\ \hline
Kelly        & 7563  & 5914    & 80.74   \\ \hline
Marlon       & 7167  & 5607    & 81.34   \\ \hline
Toucan       & 6926  & 5352    & 75.76   \\ \hline
Lion         & 7243  & 5650    & 74.81   \\ \hline
Dog          & 7891  & 6055    & 82.12   \\ \hline
\end{tabular}
\begin{tabnote}
Statistics for our results. Initial disks is our initial estimate
of the required number of disks, which serves as a starting point for a binary
search, after which we achieve the final number of disks. The percentage of
correctly sized disks among these are reported in the right column.
\end{tabnote}
\label{tab:stats}
\end{table}

The resulting lampshade designs have been 3D-printed to allow a qualitative
evaluation of the projected imagery.
The lampshades were printed on a powder-based
binder-jet printer (ProJet$^\circledR$ 660Pro). The thickness of the lampshade shells was set to 3mm.
The manufacturing time for such a spherical lampshade includes 16.5 hours
of printing and 1 hour of drying. The inside surface of the lamps were
sprayed with black paint, which somewhat improves the contrast of the
projected images.

Table~\ref{tab:stats} reports several statistics for each of these results.
The computation time of our method consists of two main parts:
computing the disk distribution and generating the 3D model of the lampshade
perforated with tubes. For the lamps shown in the paper it takes less
than 5 minutes to run the method of de Goes et al.~\shortcite{deGoes2012}
up to 10 times, while searching for the optimal number of disks. The resulting
lamps have around 6000 tubes, and the 3D model generation takes roughly
3 minutes. For verification, we typically compute a simulated result before
sending a lamp to be printed, which takes roughly 1 minute. All of the
above times are measured on an Intel$^\circledR$ Core\textsuperscript{TM} i5 CPU \@3.3GHz with 8GB RAM.

\begin{figure}[tb]
\centering
\includegraphics[width=.3\linewidth]{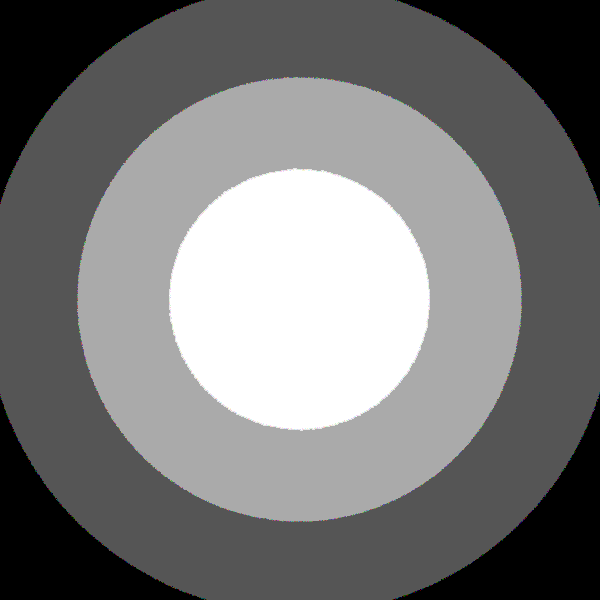}~
\includegraphics[width=.3\linewidth]{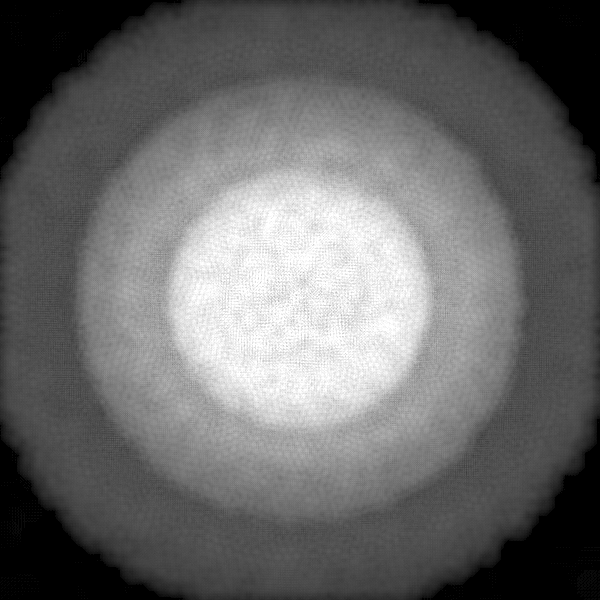}~
\includegraphics[width=.3\linewidth]{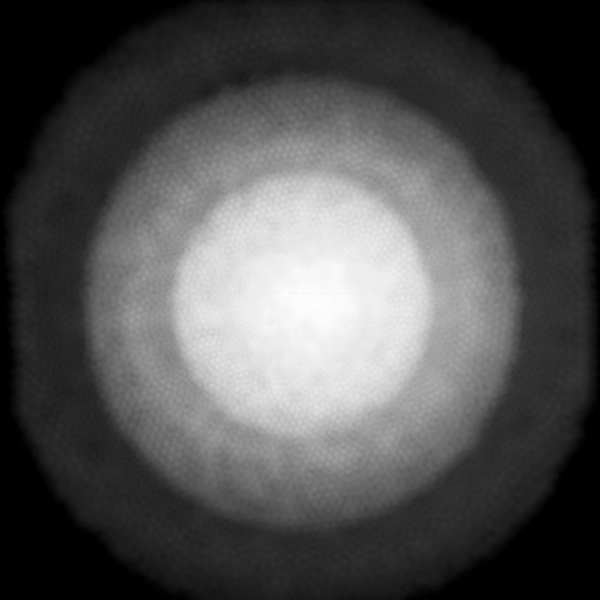}\\
\centerline{(a)\hspace{0.27\linewidth}(b)\hspace{0.27\linewidth}(c)}
\includegraphics[width=.3\linewidth]{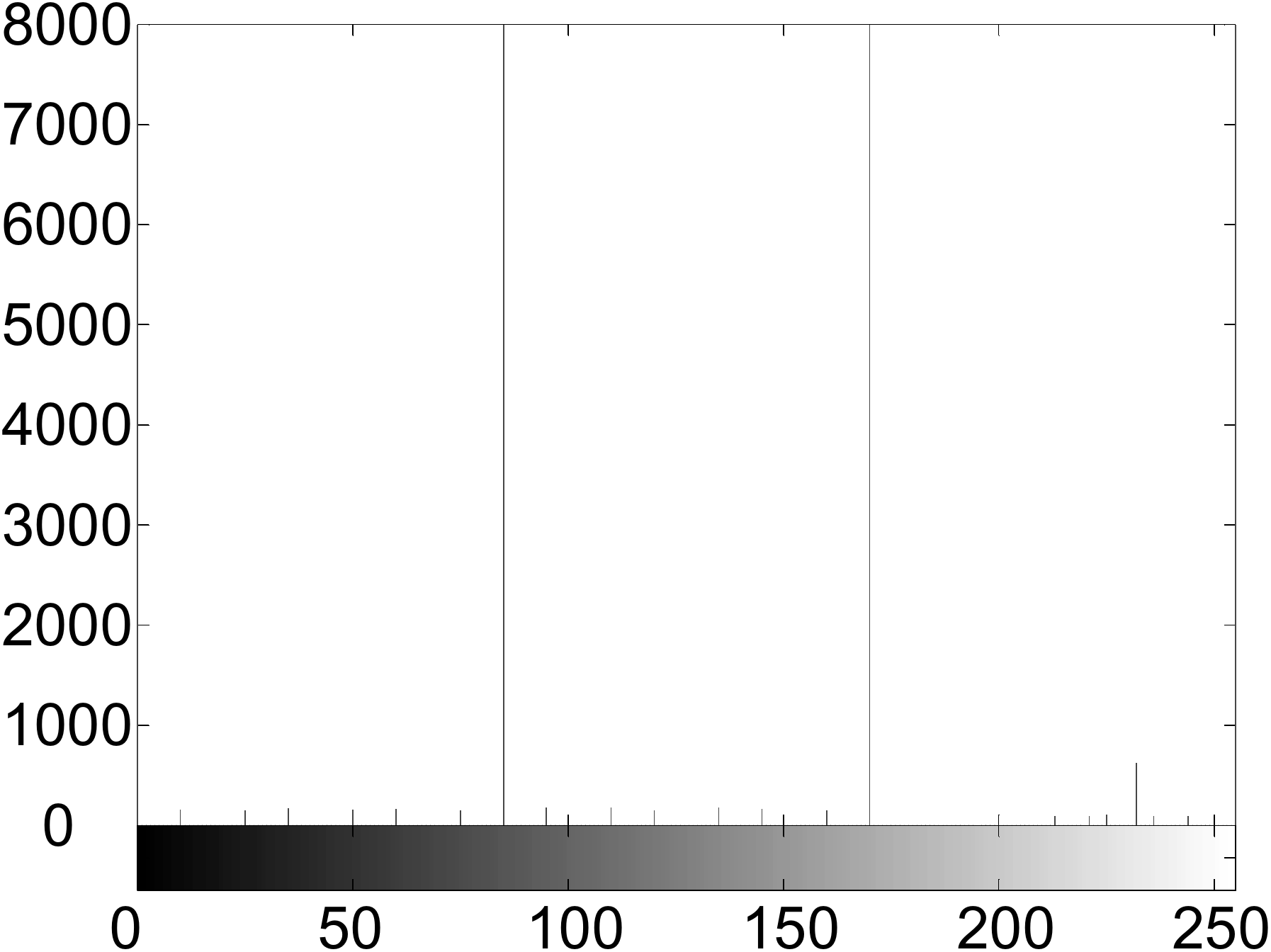}~
\includegraphics[width=.3\linewidth]{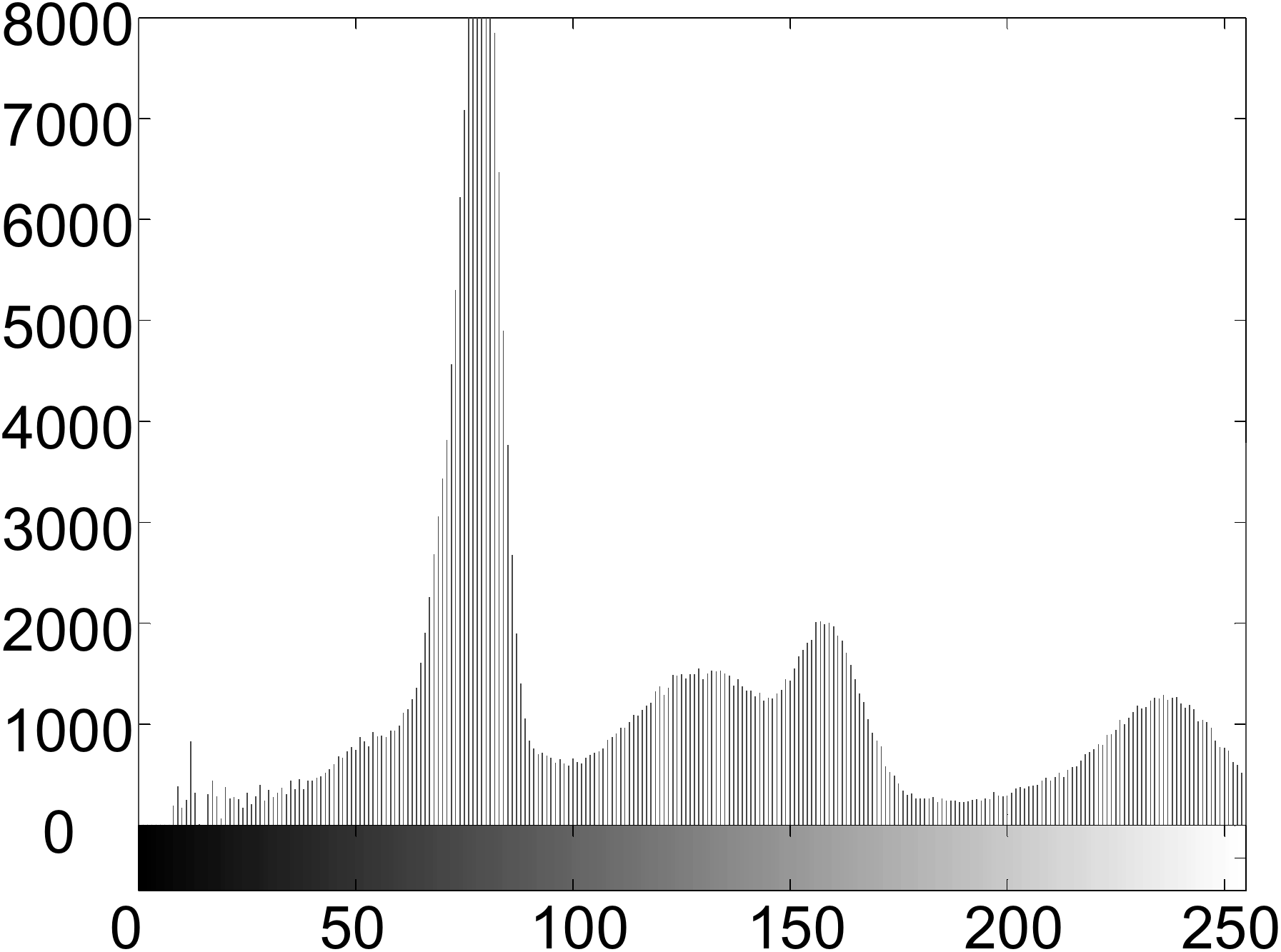}~
\includegraphics[width=.3\linewidth]{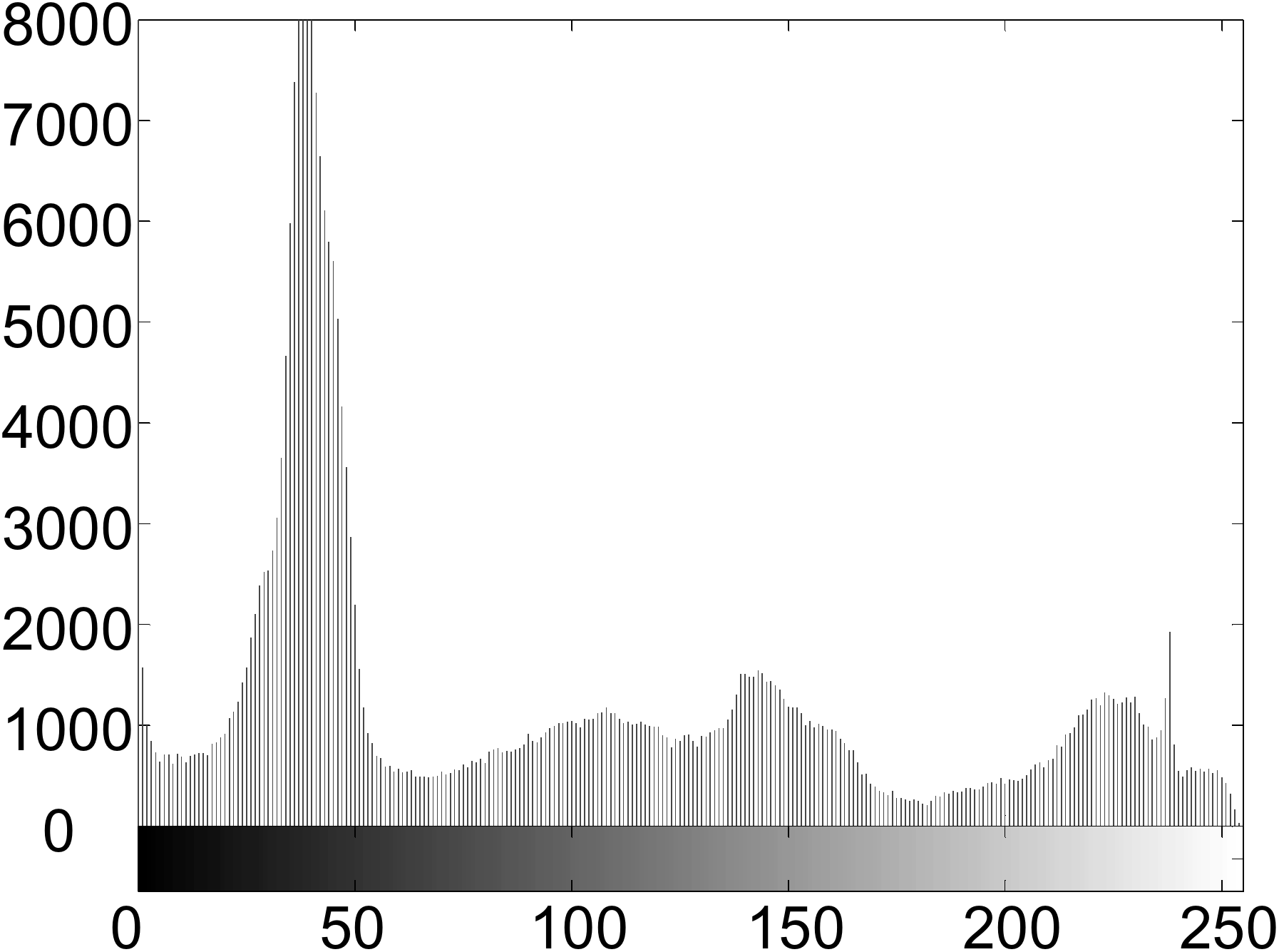}\\
\centerline{(d)\hspace{0.27\linewidth}(e)\hspace{0.27\linewidth}(f)}

\includegraphics[width=.3\linewidth]{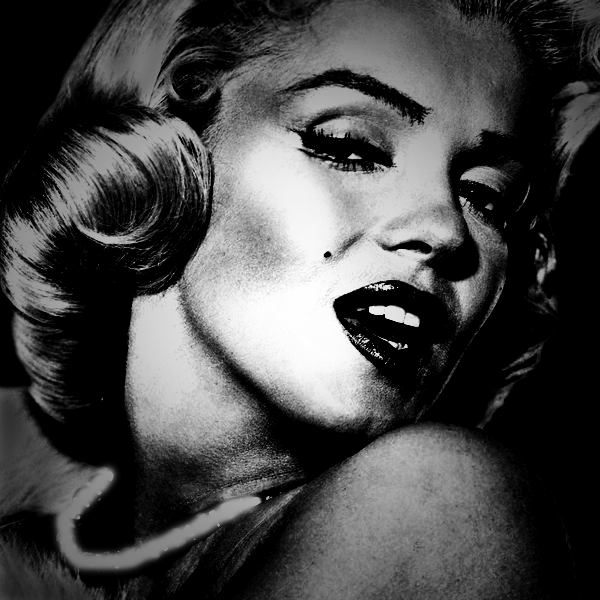}~
\includegraphics[width=.3\linewidth]{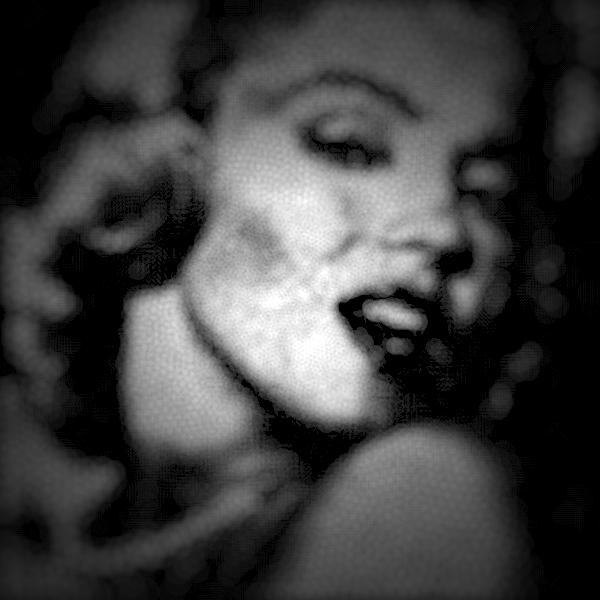}~
\includegraphics[width=.3\linewidth]{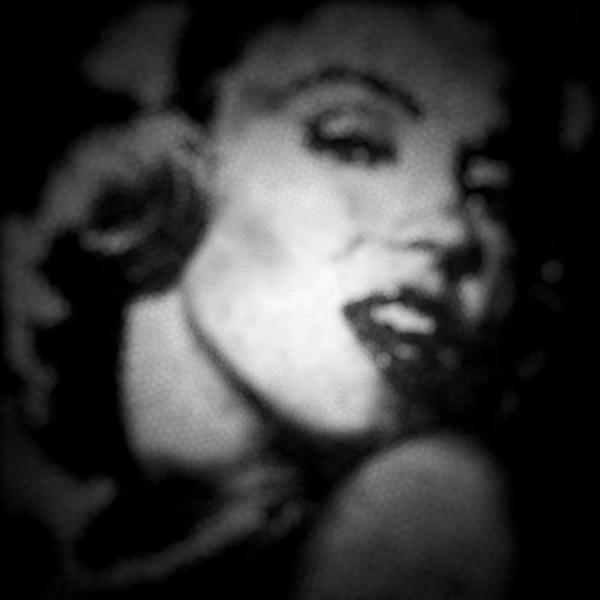}\\
\centerline{(h)\hspace{0.27\linewidth}(i)\hspace{0.27\linewidth}(j)}
\includegraphics[width=.3\linewidth]{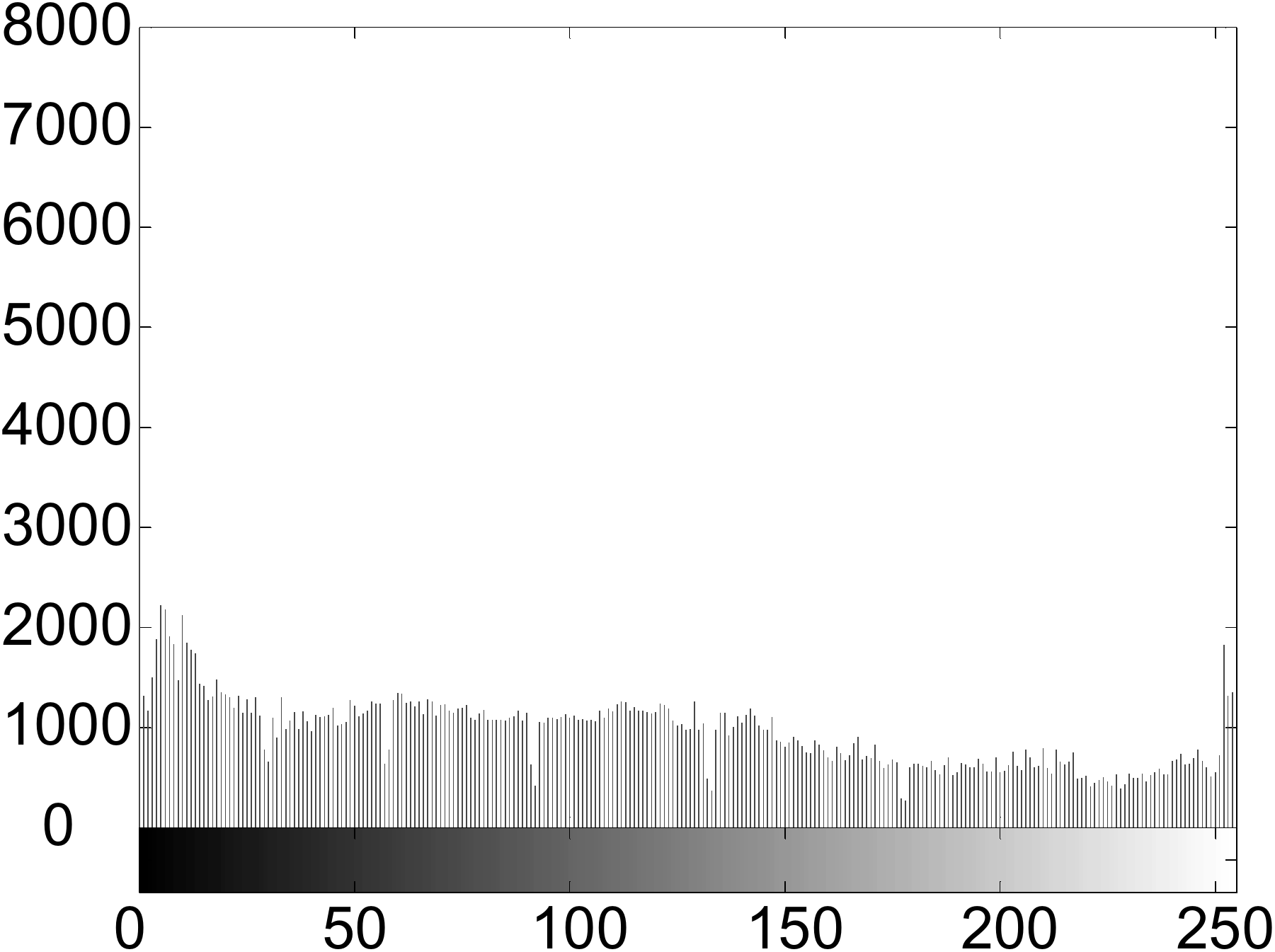}~
\includegraphics[width=.3\linewidth]{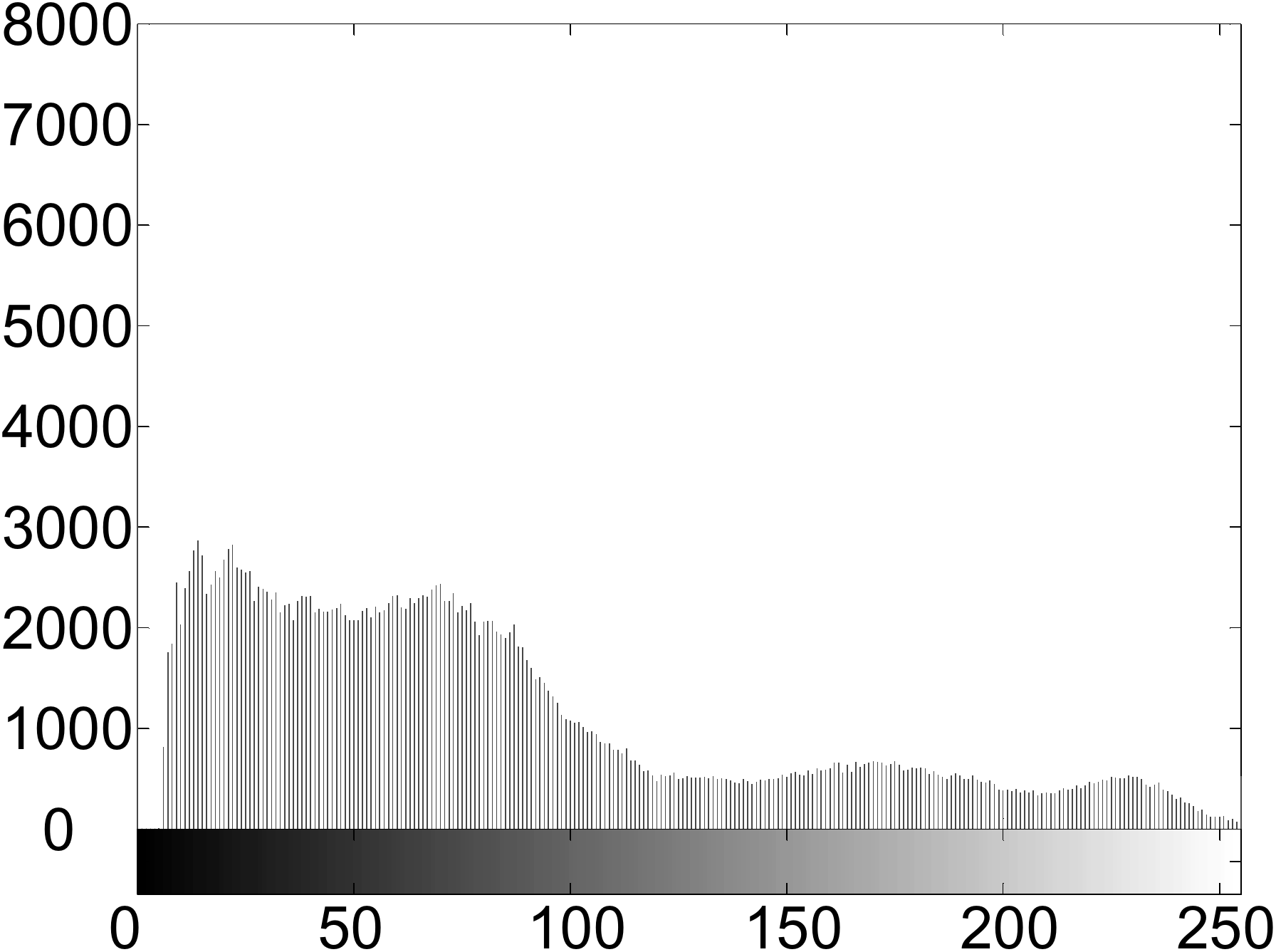}~
\includegraphics[width=.3\linewidth]{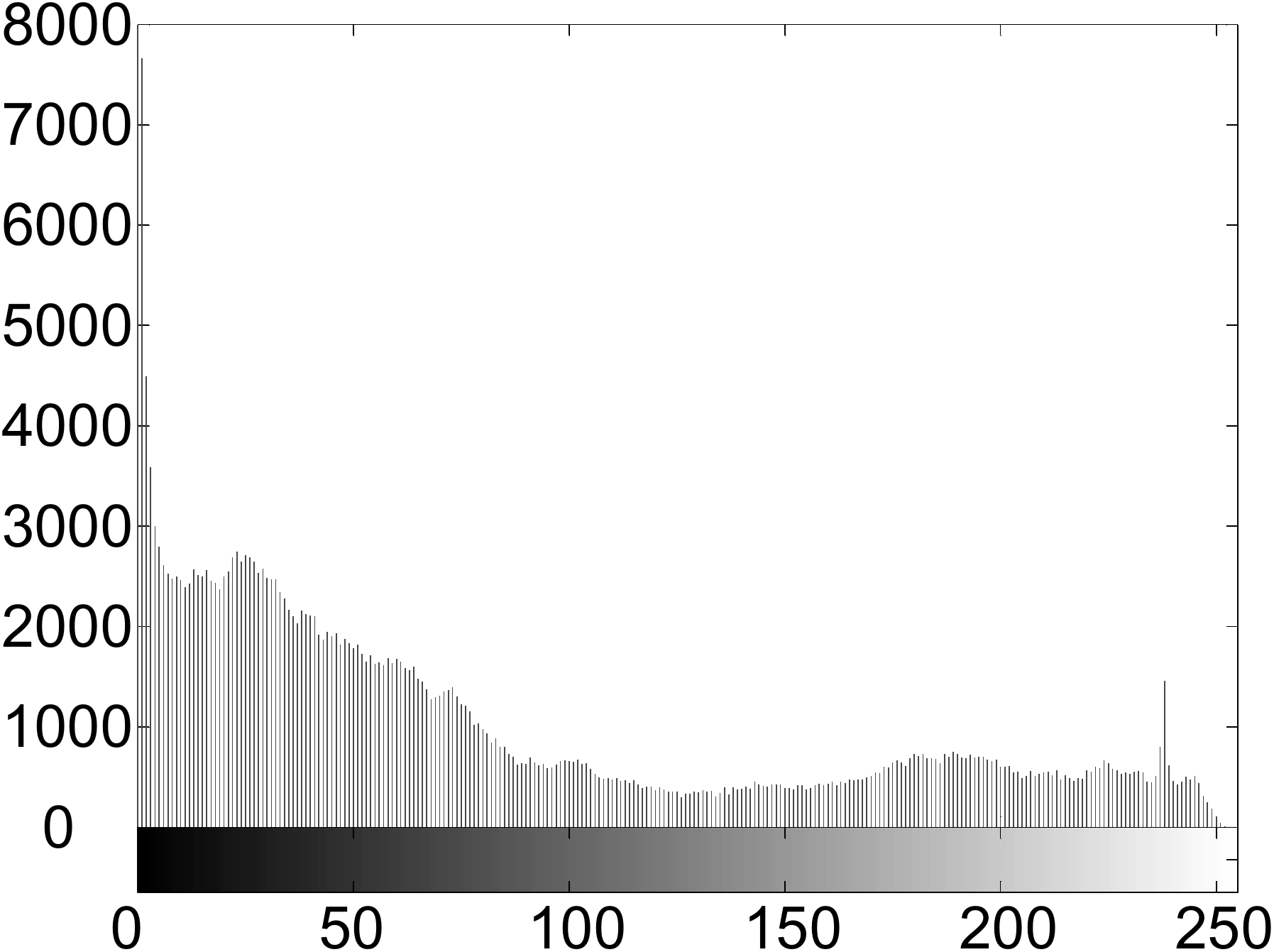}\\
\centerline{(k)\hspace{0.27\linewidth}(l)\hspace{0.27\linewidth}(m)}
\caption{
Two comparisons of the target image (a,h), our simulation result (b,i), and a real photo of the projected image (c,j).
The corresponding histogram is below each image.
The circular ramp (a) consists of three gray values: 85, 170, 255.}
\label{fig:ramp}
\end{figure}

Figure \ref{fig:ramp} shows two qualitative comparisons between the target
image, the simulated result, and the physical projected image. This is done on
two examples: a concentric circles pattern featuring four regions of a constant
grayscale value, and a more complex natural image. Below each image,
we show the corresponding histogram,
in order to better examine the differences between
the corresponding tone distributions.

In the concentric ramp case (\ref{fig:ramp}a), it obviously infeasible to
reproduce the ideal four-way distribution of grayscale values with our process.
Our design process approximates this distribution with four wider peaks
(the black peak coincides with the left edge of the histogram). A photograph
of the actual projection features four similar peaks, although there are
slight differences in their positions and spread. The target image of Marilyn
Monroe (\ref{fig:ramp}h)
has a nearly uniform histogram. This histogram is reproduced reasonably well
by the our process, however, both the simulated and the actual projected
image have more mass in the lower part of the grayscale range.

These two examples illustrate well some of limitations of this new
medium. Achieving bright tones in the middle of the image is relatively easy
(since the distance between the receiver and the light source is minimal there,
and both cosines in Eq. (\ref{eq:illum}) are near 1.0). However, reproducing
dark tones in the central region is challenging because of the necessity to use
large tilt angles. As explained earlier, extensive use of tilting requires
reducing the tube density, thereby reducing spatial resolution. Conversely,
achieving light tones at the periphery of the image is difficult, because of the
falloff in the illuminance, and increasingly wider tubes are needed to produce
a light tone in the periphery. Thus, there's a tradeoff between our ability to
reproduce the full tonal range, and our desire for better spatial
resolution. In particular, by imposing an upper bound (1.3mm) on the embedding
disk radii across the lamp surface, we are not able to match the lighter tones
of the target image beyond a certain radius.

\begin{figure}[tb]
\centering
\includegraphics[width=.16\linewidth]{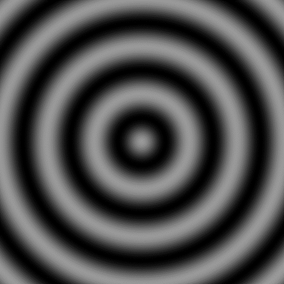}
\includegraphics[width=.16\linewidth]{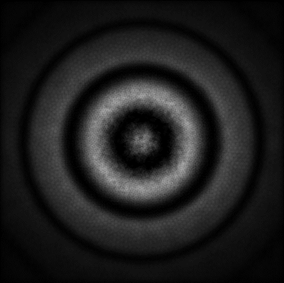}
\includegraphics[width=.63\linewidth]{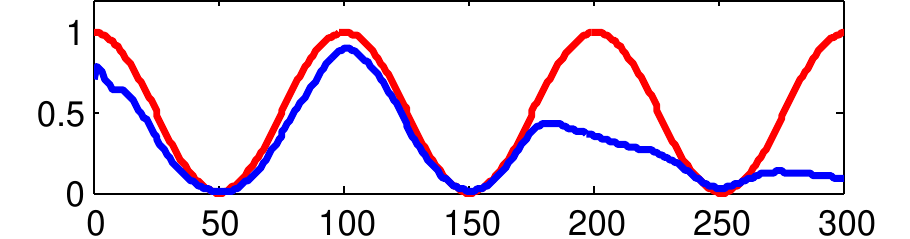}\\
\includegraphics[width=.16\linewidth]{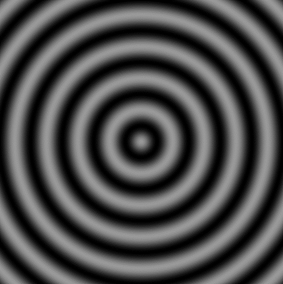}
\includegraphics[width=.16\linewidth]{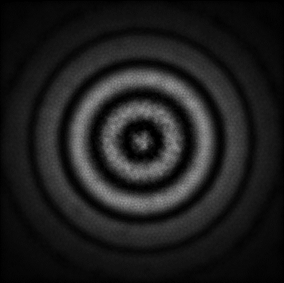}
\includegraphics[width=.63\linewidth]{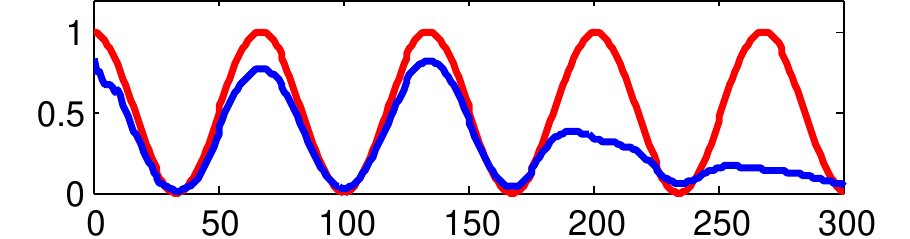}\\
\includegraphics[width=.16\linewidth]{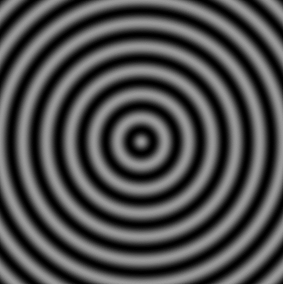}
\includegraphics[width=.16\linewidth]{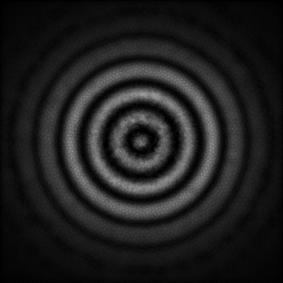}
\includegraphics[width=.63\linewidth]{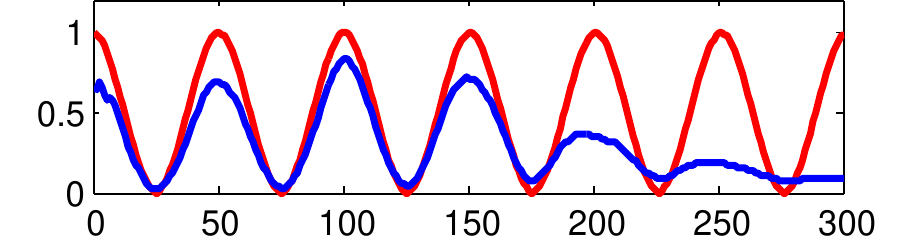}\\
\includegraphics[width=.16\linewidth]{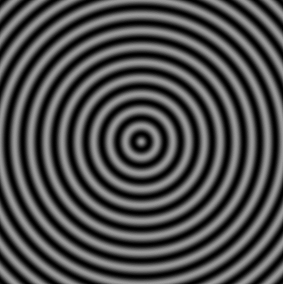}
\includegraphics[width=.16\linewidth]{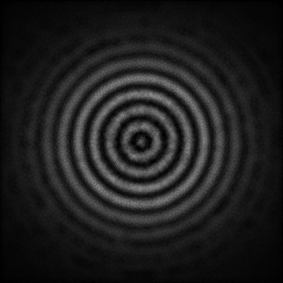}
\includegraphics[width=.63\linewidth]{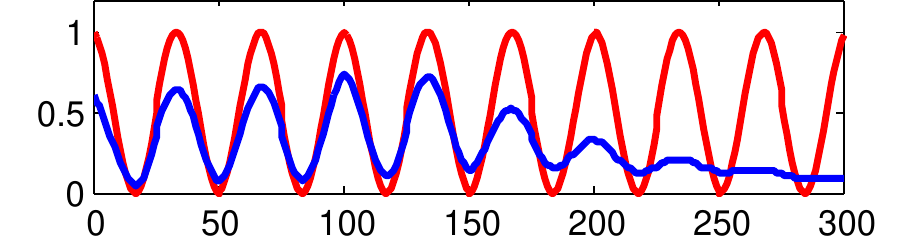}\\
\includegraphics[width=.16\linewidth]{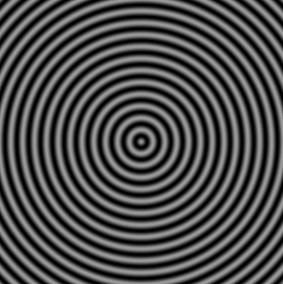}
\includegraphics[width=.16\linewidth]{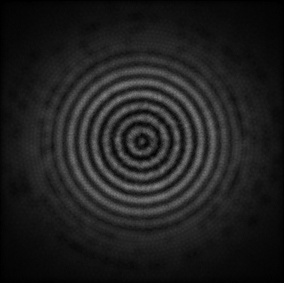}
\includegraphics[width=.63\linewidth]{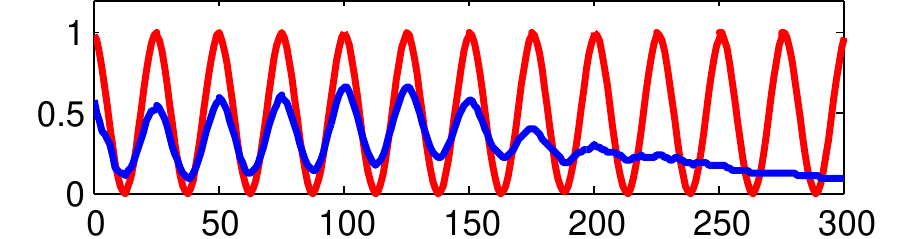}\\
\includegraphics[width=.16\linewidth]{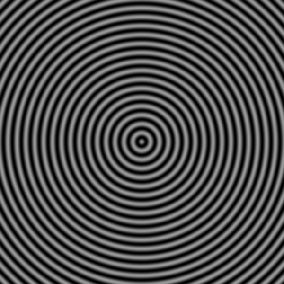}
\includegraphics[width=.16\linewidth]{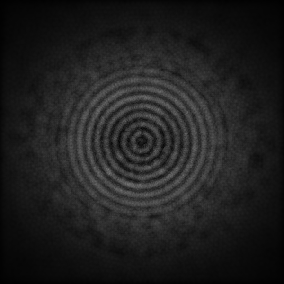}
\includegraphics[width=.63\linewidth]{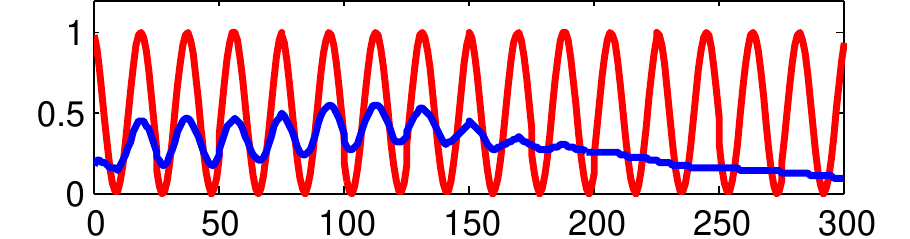}\\
\caption{Frequency and contrast response. Given the radial cosine waves as the input pattens (left), the results of our method (middle) are influenced by both frequency and radial distance from the center, as shown in the amplitude plot (right). }
\label{fig:freq}
\end{figure}

The above can be observed in a more quantitative manner in Figure \ref{fig:freq},
which shows the response of our method to several radial cosine
waves with the frequency increasing from top to bottom row. Next to each input
pattern we show the simulated result produced by our method and plot the
amplitudes of the input wave (in red) and the result (blue), as a function
of the radial distance. It may be
observed that for all frequencies our method fails to reproduce the
waveform beyond a certain radial distance, with the distance slightly
decreasing as the frequency increases. It may also be seen that in the areas
where the waveform is reproduced, the contrast response also decreases
with frequency: while in the top row the ratio between the reproduced amplitude
to the input one is roughly 0.89, this ratio drops to 0.40 in the bottom row.

\begin{figure}[ht]
\centering
\includegraphics[width=.8\linewidth]{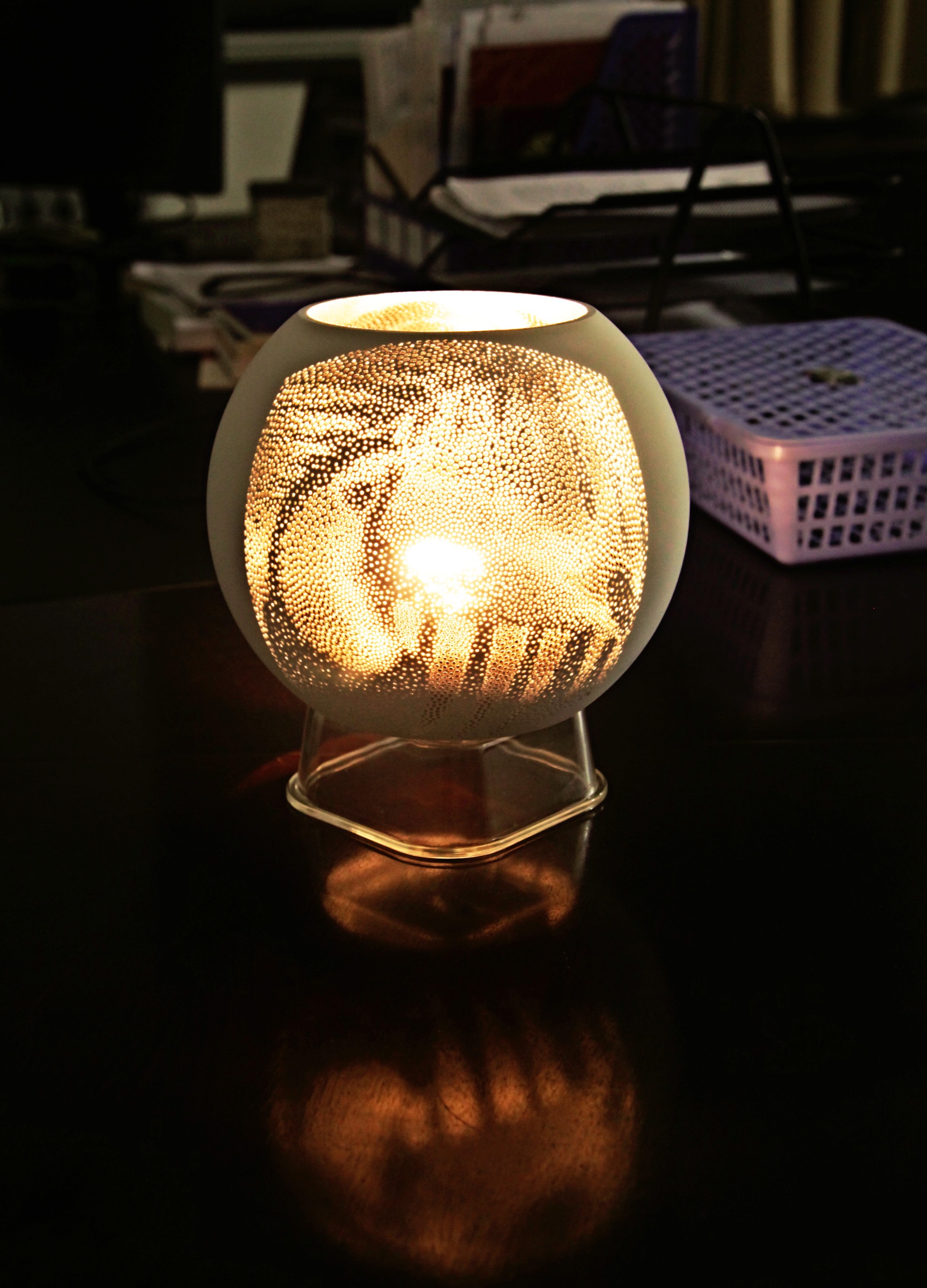}
\caption{One of our manufactured lamps in a natural setting.}
\label{fig:reallamp}
\end{figure}

%% file: conclusion.tex
\section{Concluding Remarks}
\label{sec:conclusion}

We have presented a technique to design printable perforated lampshades. The main challenge was to control the light emanating through the lampshade under the unique printability constraints and engineering setting. The technique that we developed extends classical halftoning to a novel domain, where the ``dots'' are not printed, or physically tangible, but projected, effectively \emph{halftoning with light}.

Although we have not yet adapted our process to non-spherical lampshades, note
that the process is general enough to accommodate such shapes. Specifically,
given a non-spherical lampshade geometry, one would need to (a) distribute
dense uniform patterns of tubes across the lampshade, in order to compute
the simulated reference images $B_i$; and (b) determine an appropriate
transformation between disks on the lampshade surface and corresponding
disks on the projection surface. The rest of our process would remain unchanged.

The 3D printing technology is emerging and growing quickly, and many new intriguing applications are introduced. We believe that the combination of light and 3D printing has much more to offer. One direction which we now consider is not to use the light directly, but indirectly, by using the printed surface as reflector. The printed surface can be customized to the given environment, to its geometry and photometric properties, so as to optimize the distribution of light by a
custom-made reflector.

Another direction for future work is to go beyond the printability constraints, by assembling a large scale lampshade from surface pieces printed separately. This will allow using bigger and stronger light sources, and possibly compound arrays of light sources. A larger surface area will allow increasing the relative resolution of printed holes, and project an image on larger and more distant receiving surfaces. We are also considering focusing on the design of aesthetic imagery on the lampshade itself. Halftoned and stippled images have their own aesthetic virtue. In Figure~\ref{fig:reallamp}, we show our lamp in a bedroom. Light effects combined with halftoning can lead to creative artistic media.